\documentclass{aa}  

\usepackage{graphicx}
\usepackage{xcolor}
\usepackage{txfonts}
\usepackage[colorlinks,linkcolor={blue},citecolor={blue}]{hyperref}
\usepackage{url}
\renewcommand{\d}[1]{\ensuremath{\operatorname{d}\!{#1}}}

\begin{document} 
   
   \title{Simulations of the line-driven instability in magnetic hot star winds}
     
  \titlerunning{Wind instability of magnetic hot stars}
  
   \author{F.~A.~Driessen\and N.~D.~Kee\thanks{Current address: National Solar Observatory, 22 Ohi'a Ku St., Makawao, HI 96768, USA}\and J.~O.~Sundqvist}
   \institute{Institute of Astronomy, KU Leuven,
              Celestijnenlaan 200D, 3001 Leuven, Belgium\\
              \email{florian.driessen@kuleuven.be}
             }
             
   \date{Received ....; accepted ...}

 
  \abstract
   {Line-driven winds of hot, luminous stars are intrinsically unstable due to the line-deshadowing instability (LDI). In non-magnetic hot stars, the LDI leads to the formation of an inhomogeneous wind consisting of small-scale, spatially separated clumps that can have great effects on observational diagnostics. However, for magnetic hot stars the LDI generated structures, wind dynamics, and effects on observational diagnostics have not been directly investigated so far. }   
   {We investigated the non-linear development of LDI generated structures and dynamics in a magnetic line-driven wind of a typical O-supergiant.}
   {We employed two-dimensional axisymmetric magnetohydrodynamic (MHD) simulations of the LDI using the Smooth Source Function approximation for evaluating the assumed one-dimensional line force. To facilitate the interpretation of these magnetic models, they were compared with a corresponding non-magnetic LDI simulation as well as a magnetic simulation neglecting the LDI. }
   {A central result obtained is that the wind morphology and wind clumping properties change strongly with increasing wind-magnetic confinement. Most notably, in magnetically confined flows, the LDI leads to large-scale, shellular sheets (`pancakes') that are quite distinct from the spatially separate, small-scale clumps in non-magnetic line-driven winds. We discuss the impact of these findings for observational diagnostic studies and stellar evolution models of magnetic hot stars.}
   {}


   \maketitle
%

\section{Introduction}

Modern spectropolarimetric surveys employing the Zeeman splitting in spectral lines have firmly established that hot, luminous (OB) stars can harbour global surface magnetic fields. These surface magnetic fields are mainly dipolar and have field strengths ranging from a few 100\,G to some 10\,kG. Based on the current observational detection threshold, recent surveys show that only a modest fraction ($\approx$7\%) of Galactic single and binary OB stars harbour such surface magnetic fields (\citealt{2015A&A...582A..45F} (BOB); \citealt{2015IAUS..307..330A} (BinaMIcS); \citealt{2016MNRAS.456....2W,2019MNRAS.489.5669P} (MiMeS)), while outside the Galaxy there is not enough direct empirical evidence for magnetic OB stars yet \citep{2020A&A...635A.163B}. The origin of these surface magnetic fields remains elusive, but the rarity of magnetic OB stars and the lack of short-term observed evolution of their magnetic field does not favour a dynamo mechanism as is, an example being the case of the Sun. Instead the prevailing thought is that the magnetic fields of magnetic OB stars are fossil fields stemming from earlier stellar phases \citep{1982ARA&A..20..191B,2017arXiv170510650A}. In particular, recent research shows that proto-stellar mergers might offer a tentative explanation for surface magnetic fields in hot stars \citep{2019Natur.574..211S,2020MNRAS.495.2796S}. 

The presence of a global stellar magnetic field can have important effects in governing the stellar wind of hot stars \citep{1990ApJ...365..665S,1997A&A...323..121B,2004ApJ...600.1004O}. With the seminal work of \citet[][hereafter CAK]{1975ApJ...195..157C}, it is known that OB stars drive strong line-driven winds due to the scattering of the stellar continuum radiation in a large ensemble of spectral lines and it has become the de facto standard for time-dependent wind models \citep[e.g.][]{1990ApJ...356..591B,1995ApJ...440..308C,2016MNRAS.458.2323K,2017MNRAS.467.2585E,2018MNRAS.481.5263D,2021arXiv210709675S}. 

To that end, initial time-dependent two-dimensional magnetohydrodynamic (MHD) line-driven wind models also adopted CAK theory. Specifically these models have shown that the outwards expanding stellar wind can be magnetically channelled thereby forming a region of magnetically confined material, or, a circumstellar magnetosphere \citep{2002ApJ...576..413U}. This kind of `inside-out expansion' of the wind makes stellar magnetospheres fundamentally different from planetary magnetospheres that are due to `outside-in compression' from the solar wind. Moreover, with the help of the so-called magnetic confinement--rotation diagram \citep{2013MNRAS.429..398P}, it is possible to classify OB star magnetospheres based on magnetic and rotational properties. This leads to a distinction between fast rotating, strongly magnetic hot stars with centrifugal magnetospheres \citep[CM;][]{2005MNRAS.357..251T} and slowly rotating, moderately magnetic hot stars with dynamical magnetospheres \citep[DM;][]{2012MNRAS.423L..21S}. 

The use of CAK theory together with state-of-the-art spectropolarimetric observations have thus played a major role in enhancing the understanding of circumstellar magnetospheres of OB stars. However, a key limitation of these models is that they use CAK theory, which relies on the Sobolev approximation to compute the radiative acceleration of the wind \citep{1984ApJ...282..591F,2002ApJ...576..413U,2005MNRAS.357..251T,2007MNRAS.382..139T,2008MNRAS.385...97U,2009MNRAS.392.1022U,2016MNRAS.462.3672B,2017AN....338..868K,2019MNRAS.489.3251D}. In the Sobolev approximation it is assumed that the interaction of photons and spectral lines is described in a purely local fashion \citep{1960mes..book.....S} with the important corollary that the stellar wind remains smooth and homogeneous and only external effects can break the smoothness (e.g.~magnetic fields or rapid rotation).

In fact, it is nowadays well known that line-driven winds of non-magnetic hot stars are intrinsically inhomogeneous. The Doppler effect enhances small-scale velocity perturbations and exposes a spectral line to fresh, unattenuated continuum photons thereby leading to a runaway effect, or line-deshadowing instability (LDI) \citep{1979ApJ...231..514M,1980ApJ...241.1131C,1984ApJ...284..337O}. The LDI, however, is triggered on small spatial scales not covered within the Sobolev approximation. Time-dependent radiation-hydrodynamic simulations of the LDI show that once this instability is triggered a vigorous growth occurs leading to large non-linearities. The transition from the (microscopic) linear to the (macroscopic) non-linear regime is very subtle and the LDI is subject to wave-stretching \citep{2017MNRAS.469.3102F}. Once in the non-linear regime the characteristic wind morphology becomes that of slow, overdense, clumpy structures embedded in a fast, rarefied medium \citep{1988ApJ...335..914O,1995A&A...299..523F,2002A&A...381.1015R,2005A&A...437..657D,2018A&A...611A..17S,2019A&A...631A.172D,2021A&A...648A..94L}. The presence of such wind clumps is able to explain several observational phenomenae across the electromagnetic spectrum \citep[][for a review]{2008A&ARv..16..209P}, and, especially, the clumps have a range of important effects on the interpretation of observational diagnostics \citep[e.g.][]{2015IAUS..307...25P}.

For magnetic line-driven winds the effects of the LDI are poorly understood both from a theoretical and observational point-of-view. To date the only attempt to theoretically study the LDI in a magnetic line-driven wind is \citet{2020MNRAS.499.4282D}, who performed a three-dimensional linear stability analysis of magneto-radiative waves in a magnetic line-driven wind. A key insight from this study is that scattered photons provide a strong damping mechanism for short-wavelength, radially propagating magnetic waves in the wind. The exact effects of this damping on the non-linear wind dynamics and its potential observational signature remain to be investigated and necessarily require multi-dimensional radiation-MHD simulations.

In this paper we present the first effort in performing two-dimensional radiation-MHD simulations of the LDI. In particular, Sect.~\ref{sec:hydro} presents the underlying computational details we have employed in our modelling. Our results regarding wind morphology and statistical properties are discussed within Sect.~\ref{sec:results}. We also briefly discuss the potential observational signatures of the LDI in a magnetic line-driven wind and frame our results with respect to stellar evolution models of magnetic hot stars (Sect.~\ref{sec:obs}). Finally, Sect.~\ref{sec:conc} summarises our findings and suggests directions for further research. To gain further insight, Appendix \ref{sec:comparison} briefly compares our magnetic LDI wind models with previous wind models employing the CAK line-force parametrisation for the radiative acceleration.

\section{Modelling of magnetic line-driven winds}\label{sec:hydro}

\subsection{Magnetic characterisation and adopted models}

The important competition between the stellar wind and the magnetic field can be conveniently described by comparing their energies, that is wind kinetic energy vs.~magnetic energy, and is quantified with a dimensionless quantity
\begin{equation}\label{eq:eta}
\eta_\star = \frac{B_\star^2 R_\star^2}{\dot{M}_{B=0} \varv_\infty},
\end{equation}
called the wind-magnetic confinement parameter \citep{2002ApJ...576..413U}. In terms of this parameter, $\eta_\star \gg 1$ means a dominating magnetic field over the wind flow at the stellar surface and $\eta_\star \ll 1$ vice versa. It is set by the stellar magnetic field strength in the equator $B_\star = B_p/2$ in terms of the polar magnetic field strength $B_p$ and the wind terminal momentum given by a mass-feeding rate $\dot{M}_{B=0}$ (assumed to be the mass-loss rate of an equivalent non-magnetic star) at terminal wind speed $\varv_\infty$.

If confinement occurs this is over a constricted spatial extent since the magnetic energy always falls off faster than the wind energy \citep[e.g.~for observed dipole fields $B \propto 1/r^3$,][]{2017MNRAS.465.2432G}. For a prototypical line-driven wind the spatial point where the wind starts to dominate is about
\begin{equation}\label{eq:ralf}
R_A/R_\star \approx 1 + (0.25 + \eta_\star)^{1/4} - 0.25^{1/4},
\end{equation}
also known as the Alfv\'en radius \citep{2002ApJ...576..413U}. Because we do not consider rotation (see below), the Alfv\'en radius is the only characteristic magnetosphere scale in the present work. This means that the magnetospheres here are Dynamical Magnetospheres (DM). Consulting the magnetic confinement--rotation diagram these are intrinsic to slowly rotating magnetic O-stars \citep{2013MNRAS.429..398P}. Therefore, for our modelling we adopt stellar and wind parameters typical to a O-supergiant in the Galaxy as collected in Table\,\ref{table:simparams}. 

Since the present work is concerned with a first study of a more complete description of the radiation line force in a magnetic line-driven wind this warrants some simplifications for other physics. We assume an isothermal wind fixed at the stellar effective temperature, that is heating from shocks is radiated away in an unresolved small cooling layer \citep{1997A&A...322..878F,2021A&A...648A..94L}. Similarly, we do not consider additional effects such as (rapid) stellar rotation \citep{2006ApJ...640L.191U,2008MNRAS.385...97U}, or an oblique dipolar magnetic field topology \citep{2019MNRAS.489.3251D} that necessarily require a three-dimensional model.

\subsection{Magnetohydrodynamics}
 
The employed conservative single-fluid, ideal magnetohydrodynamic (MHD) equations solve for the mass density $\rho$, velocity field $\mathbf{v}$, and magnetic field $\mathbf{B}$
\begin{equation}\label{eq:mass}
\frac{\partial \rho}{\partial t} + \nabla \cdot (\rho \mathbf{v}) = 0,
\end{equation}
\begin{equation}\label{eq:mom}
\frac{\partial (\rho\mathbf{v})}{\partial t} + \nabla \cdot \left[ \rho\mathbf{v}\mathbf{v} + \left( p + \frac{\mathbf{B}\cdot \mathbf{B}}{2}\right)\mathbf{I} - \mathbf{B}\mathbf{B} \right] = \rho\mathbf{g}_\mathrm{eff}+ \rho\mathbf{g}_\mathrm{line},
\end{equation}
\begin{equation}\label{eq:bf}
\frac{\partial \mathbf{B}}{\partial t} + \nabla \cdot (\mathbf{v}\mathbf{B} - \mathbf{B}\mathbf{v}) = 0.
\end{equation}
These are complemented with the divergence-free constraint
\begin{equation}\label{eq:divb}
\nabla \cdot \mathbf{B} = 0,
\end{equation}
and isothermal closure relation for the gas pressure
\begin{equation}
p = a^2 \rho, \qquad a = \sqrt{k_B T/m},
\end{equation}
with $a$ the isothermal speed of sound defined from the stellar effective temperature, namely from assuming $T_\mathrm{wind}=T_\mathrm{eff}$, and Boltzmann's constant $k_B$ for a gas with mean atomic weight $m$. We point out that in this formulation of the (dimensionless) equations the magnetic field is defined such that the magnetic permeability is $\mu_0=1$. 

Additional source terms are included in the momentum equation to account for stellar gravity and the radiation force consisting of continuum and line interactions. In hot star winds the continuum force is primarily due to electron scattering, which effectively lowers gravity, such that it is customary to formulate both quantities together into an effective stellar gravity
\begin{equation}
\mathbf{g}_\mathrm{eff} = -\frac{G M_\star (1-\Gamma_e)}{r^2}\, \mathbf{\hat{e}_r},
\end{equation}
with $\Gamma_e \equiv \kappa_e L_\star/(4\pi G M_\star c)$ the Eddington factor for constant electron scattering with opacity $\kappa_e =0.34$\,cm$^2$\,g$^{-1}$ (assuming full ionisation at solar abundances). The force due to lines, $\mathbf{g}_\mathrm{line}$, is the important quantity in modelling a line-driven wind and will be detailed next.

\subsection{Radiation line force}\label{sec:radforce}

A key difference with previous work on magnetic line-driven winds is that we depart from the line force description $\mathbf{g}_\mathrm{line}$ within the Sobolev approximation (CAK, but see Appendix \ref{sec:comparison} for a comparison). However, with the added complexity of the line force description we still opt to treat the radiation transport by calculating it along an isolated one-dimensional ray within each colatitudinal cone of the simulation domain. In that respect our approach is the magnetic equivalent of the hydrodynamic models performed by \citet{2003A&A...406L...1D}. 

The total line force (per unit mass) consists of direct and diffuse contributions due to photon absorption and scattering: $g_\mathrm{line} \equiv g_\mathrm{dir} + g_\mathrm{diff}$. In driving the wind many lines participate and the cumulative contribution of all lines is described by an ensemble power-law distribution (CAK) in line strength $q$ \citep{1995ApJ...454..410G} with an additional exponential cut-off \citep{1988ApJ...335..914O} limiting the distribution to a line with maximum strength $Q_\mathrm{max}$: $N(q) \propto q^{\alpha-2} \exp{(-q/Q_{\mathrm{max}})}$, with $\alpha$ the power-law index of the assumed line distribution.

Within the non-Sobolev, Smooth Source Function approach \citep[SSF,][]{1996ApJ...462..894O} the ensemble-integrated direct and diffuse forces evaluate to
\begin{equation}\label{eq:gdir}
g_{\mathrm{dir}}(r) = g_{\mathrm{thin}} \int_0^1 b_+(\mu_y,r)\, \d{y},
\end{equation}
\begin{equation}\label{eq:gdiff}
g_{\mathrm{diff}}(r) = g_{\mathrm{thin}} \frac{S(r)}{I_\star} \left( \frac{r}{R_\star} \right)^2 \int_0^1 \left( b_-(\mu_y,r) - b_+(\mu_y,r) \right)\, \d{y},
\end{equation}
with optically thin line force and optically thin source function
\begin{equation}
g_\mathrm{thin} \equiv \frac{\bar{Q}\kappa_e L_\star}{4\pi r^2 c}, \qquad S(r) \equiv I_\star \frac{(1-\mu_\star)}{2},
\end{equation}
and $\bar{Q}$ is a line-strength normalisation describing the ratio of the total line force to the force due to electron scattering if all lines were to be optically thin \citep{1995ApJ...454..410G}.

Formally the angle quadrature in Eq.~\eqref{eq:gdir} is to be computed over a bundle of radiation rays that intercept the stellar disc at $\sqrt{y}R_\star$ (thereby covering the full stellar disc) with direction cosine $\mu_y = \sqrt{1-y(R_\star/r)^2}$\,relative to the local direction at radius $r$. Such an approach is, however, computationally expensive and, instead, we only apply radial radiation rays within each separate latitudinal cell. Nonetheless, this single-ray quadrature is not truly radial since each ray is additionally weighted with $y=0.5$ to mimic the finite extent of the stellar disc \citep{2013MNRAS.428.1837S}.\footnote{The weight also introduces a fiducial tilt for the (still) radially solved rays to ensure that the wind has not infinitely many critical points which does happen for pure radial rays (CAK).} To avoid further computational complexity, the single-ray quadrature is also assumed for the diffuse radiation, Eq.~\eqref{eq:gdiff}, in a two-stream approximation for outward (`$+$') and inward (`$-$') directed radiation.

Along the ray the line-force components are computed using an ensemble-escape probability
\begin{equation}\label{eq:escprob}
b_\pm(\mu_y,r) = \Gamma(\alpha)^{1/(1-\alpha)} \int_0^{+\infty} \d{x}\, \frac{\phi \left( x-\mu_y \varv(r)/\varv_{\mathrm{th}} \right)}{\left( \bar{Q}t_\pm(x,r) + \bar{Q}/Q_{\mathrm{max}} \right)^\alpha}
\end{equation}
that takes into account the accumulated optical depth 
\begin{equation}
\begin{split}
qt_+(x,r) &= \int_{R_\star}^r  \kappa \rho(r') \phi \left( x-\mu_y \varv(r')/\varv_\mathrm{th} \right) \d{r'}, \\
qt_-(x,r) &= \int_r^{+\infty}  \kappa \rho(r') \phi \left( x-\mu_y \varv(r')/\varv_\mathrm{th} \right) \d{r'},
\end{split}
\end{equation}
with $\Gamma(\alpha)$ the Gamma function, $\kappa$ a frequency-integrated line opacity, and $\phi$ a Gaussian line-profile function at observer's frame frequency $x\equiv (\nu/\nu_0 -1) c/\varv_\mathrm{th}$. With this definition, $\bar{Q}t_\pm$ in Eq.~\eqref{eq:escprob} denotes the optical depth of a line with strength $\bar{Q}$. To capture the resonance zones in the wind acceleration region these optical depth integrals are solved by analytic integration between discrete grid points using linear interpolation in density and velocity \citep[see][their Eqs.~7-8 for a similar approach]{2017MNRAS.469.3102F}.

\begin{table}
\caption{Overview of parameters used in this work.}             
{\small{

\label{table:simparams}      
\centering                          
\begin{tabular}{l c c}       
\hline\hline                 
Name & Parameter & Value  \\  
\hline                        
   Stellar luminosity                    & $L_\star$ & $8\times 10^5\,L_\odot$   \\    
   Stellar mass                            & $M_\star$ & $50\,M_\odot$ \\
   Stellar radius                           & $R_\star$  & $20\,R_\odot$   \\
   Stellar effective temperature & $T_\mathrm{eff}$ & $40\,000$\,K    \\
   Eddington factor                     & $\Gamma_e$  & 0.42 \\
   Initial mass-feeding rate        & $\dot{M}_{B=0}$ & $2\times 10^{-6}\,M_\odot\,\mathrm{yr}^{-1}$\\
   Initial terminal wind speed     & $\varv_\infty$ & 1880\,km\,s$^{-1}$ \\
   Stellar boundary density        & $\rho_0$ & $1.2\times 10^{-11}$\,g\,cm$^{-3}$ \\
   CAK exponent                         & $\alpha$ & $0.65$ \\
   Line-strength normalisation  & $\bar{Q}$ & $2000$ \\
   Line-strength cut-off             & $Q_\mathrm{max}$ & $0.004\bar{Q}$\\
   Thermal to sound speed  ratio & $\varv_\mathrm{th}/a$ & 0.28\tablefootmark{a} \\
   Isothermal sound speed         & $a$ & 23.3\,km\,s$^{-1}$\\
   Wind-magnetic confinement & $\eta_\star$ & $0$, $0.15$\tablefootmark{b}, $15$\tablefootmark{c}\\                                   
\hline                                   
\end{tabular}
\tablefoot{\tablefoottext{a}{\citet{1990ApJ...358..199P}.} For fixed stellar and wind parameters the corresponding polar magnetic field strength is \tablefoottext{b}{$B_p\approx 100\,\mathrm{G}$}and \tablefoottext{c}{$B_p\approx 1000\,\mathrm{G}$ by using the definition of $\eta_\star$.}}

}}
\end{table}
 
\subsection{Numerical specifications}
 
To perform non-linear simulations of magnetic LDI winds we use the open-source\footnote{\url{http://amrvac.org/}; version 2.2 (Nov.~2019) in this work.}, parallel, grid-adaptive astrophysical (M)HD code \textsc{mpi-amrvac} \citep{2018ApJS..234...30X,KEPPENS2021316}. Specifically, Eqs.\,\eqref{eq:mass}--\eqref{eq:divb} are solved in an unsplit fashion using a HLL Riemann solver \citep{HLL} complemented with a parabolic spatial reconstruction scheme \citep[PPM,][]{1984JCoPh..54..174C} and third-order accurate total-variation-diminishing Runge--Kutta temporal discretisation \citep{1998MaCom..67...73G}. For stability the time step is taken as the minimum of a fixed Courant time set to be $0.3$ and an additionally computed time step due to the radiation and gravity force in Eq.\,\eqref{eq:mom} \citep[see Eq.~(8) in][]{2021A&A...648A..94L}. 
 
\subsubsection{Simulation grid}
 
The simulation spans the two-dimensional meridional $(r,\theta)$ plane and covers a radial extent $r\in [R_\star, 6R_\star]$ and colatitudinal extent $\theta \in [0, \pi]$. To resolve the steep wind acceleration region near the stellar surface, we stretch the radial grid away from the stellar surface with a factor $\Delta r_{i+1}/\Delta r_i = 1.002$ between subsequent radial zones $i$ and $i+1$. In colatitude we assume uniform spacing $\Delta \theta_j$.

This grid choice ensures that the Alfv\'en radius is sufficiently far from the outer boundary while the stretching allows us to resolve the barometric scale height of the photosphere, $a^2R_\star^2/(GM_\star(1-\Gamma_e)) \approx 0.002R_\star$, and the small spatial scales over which the instability operates. The latter implies resolving the radial Sobolev length $L_{\mathrm{Sob},r} \equiv \varv_\mathrm{th}/(\d \varv_r/\d r) \approx (\varv_\mathrm{th}/\varv) r \approx 0.01R_\star$, for which we adopt $n_r=1000$ radial zones. Similarly, we also resolve a characteristic lateral Sobolev length, $L_{\mathrm{Sob},\theta} = r \varv_\mathrm{th}/\varv_r \approx 0.5^\circ$, by using $n_\theta=384$ zones in colatitude. As demonstrated by linear stability analysis, the flow should be stable on scales smaller than the lateral Sobolev length \citep{2020MNRAS.499.4282D}.
 
\subsubsection{Initial and boundary conditions}

We initialise the line-driven wind model by starting from a relaxed one-dimensional spherically symmetric CAK wind that is distributed over the meridional plane. The poloidal velocity $\varv_\theta(r,\theta) =0$ initially. The magnetic field is taken to be a pure dipole with polar field strength $B_p$,
\begin{equation}\label{eq:initbfield}
B_r(r,\theta) = B_p \left( \frac{R_\star}{r} \right)^3 \cos \theta, \qquad B_\theta(r,\theta) = \frac{B_p}{2} \left( \frac{R_\star}{r} \right)^3 \sin \theta.
\end{equation}

The outer (supersonic, super-Alfv\'enic) radial boundary assumes that all hydrodynamic variables are continuous. At the lower boundary (stellar surface) the density $\rho_0$ is fixed at five times the sonic point density in order to have a subsonic flow with a typical one-dimensional non-magnetic CAK mass flux. The radial velocity $\varv_r$ is extrapolated into the first ghost cell, allowing the wind to dynamically adjust while the remaining ghost cell values of $\varv_r$ are set equal to the first ghost cell. The radial magnetic field $B_r$ is fixed in the interior to enforce a dipolar field at the surface. The poloidal magnetic field $B_\theta$ is extrapolated into the ghost cells and the flow is forced along the magnetic field by fixing $\varv_\theta$. While the simulation adapts to its steady-state base outflow, we also prohibit the velocity to become supersonic at the lower boundary. We refer the reader to Appendix \ref{sec:appendix} for an extended account on the adopted boundary conditions.

Since the simulated meridional plane contains both polar axes ($\theta =0$ and $\theta=\pi$) we employ $\pi$--boundary conditions in $\theta$. Essentially this enforces that the fluid quantities are translated an amount $\pi$ around the (singular) pole and vector quantities transform accordingly to allow for flow across the pole. This avoids numerical problems related to ill-defined geometric factors in divergence computations.
  
\subsubsection{Treating the magnetic field and the monopole condition}
 
The interaction of the LDI with the magnetic field can result in strong spatial gradients in the wind due to shock-dominated interactions. This often leads to numerical difficulties in solving the MHD equations, especially if the total magnetic field is solved for as the dependent variable. In order to circumvent such problems, we follow the method of \citet{1994JCoPh.111..381T}, available in \textsc{mpi-amrvac}, and split the total dipole magnetic field $\mathbf{B} = \mathbf{B}_0 + \delta\mathbf{B}$ in an intrinsic background dipole field $\mathbf{B}_0$ and a deviating field $\delta \mathbf{B}$. This splitting off the total magnetic field $\mathbf{B}$ and only solving for the deviated field $\delta\mathbf{B}$ is more accurate and robust in regions of strong spatial gradients. Notice that here the background field is also potential-free $(\nabla \times \mathbf{B}_0 = 0)$ and time-independent (although the method also works for time-dependent $\mathbf{B}_0$).

To make the magnetic field satisfy the divergence-free condition, Eq.\,\eqref{eq:divb}, we apply the eight-wave scheme \citep{1999JCoPh.154..284P}. Within this method any magnetic monopoles that arise in the simulation are advected away at the fluid velocity.
   
\section{MHD simulation results}\label{sec:results}

\subsection{General wind properties}\label{sec:rholdi}

\begin{figure*}[p]
\centering
\includegraphics[width=0.85\hsize]{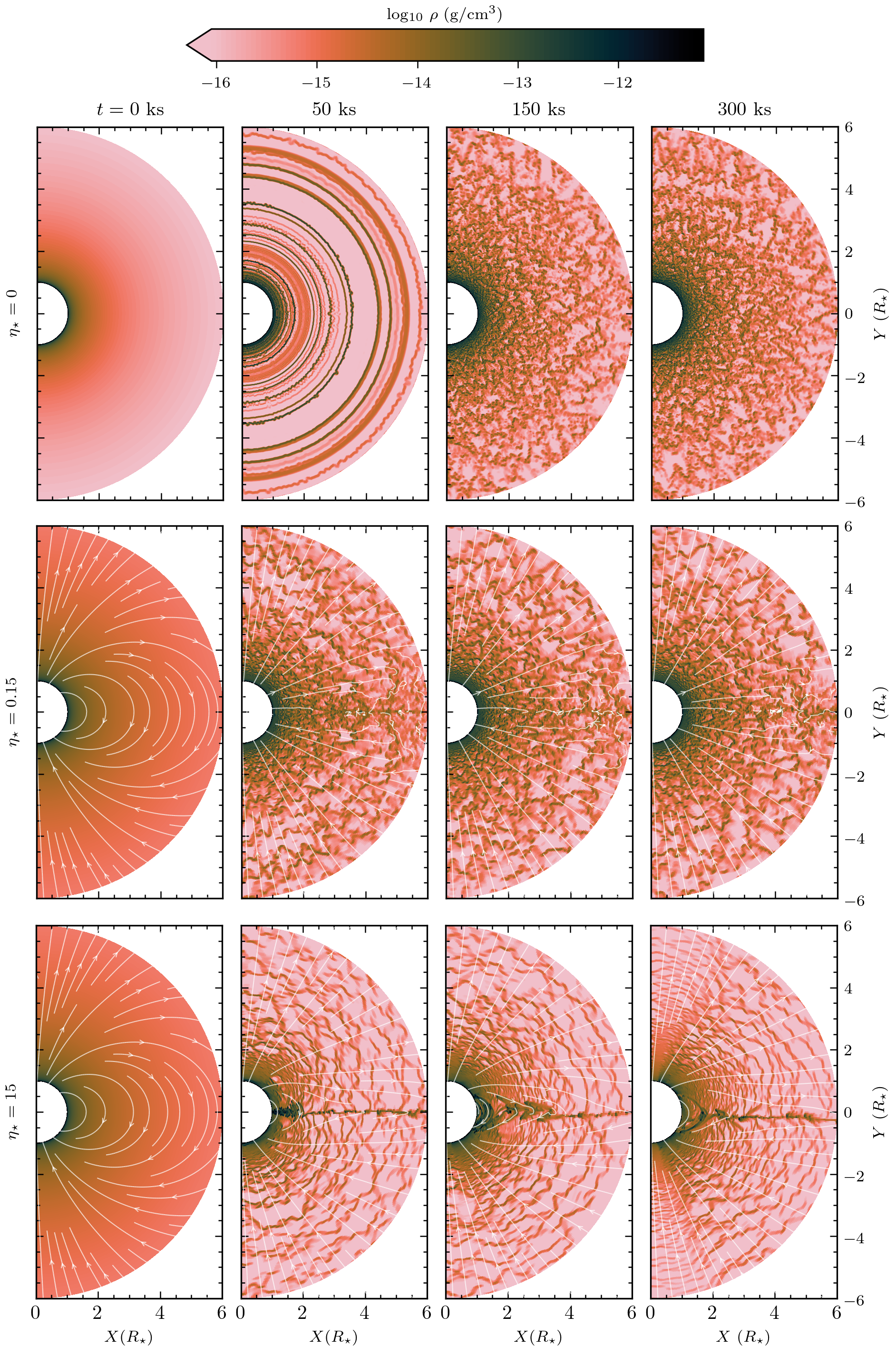}
\caption{Logarithmic wind density displaying the evolution in time for our adopted LDI models: (top) non-magnetic wind with $\eta_\star=0$, (middle) moderately confined wind with $\eta_\star=0.15$, and (bottom) strongly confined wind with $\eta_\star=15$. The overplotted white lines in the magnetic wind models represent streamlines to illustrate the magnetic field topology.}
\label{fig:rhodyn}
\end{figure*}

\begin{figure*}[ht]
\centering
\includegraphics[width=\hsize]{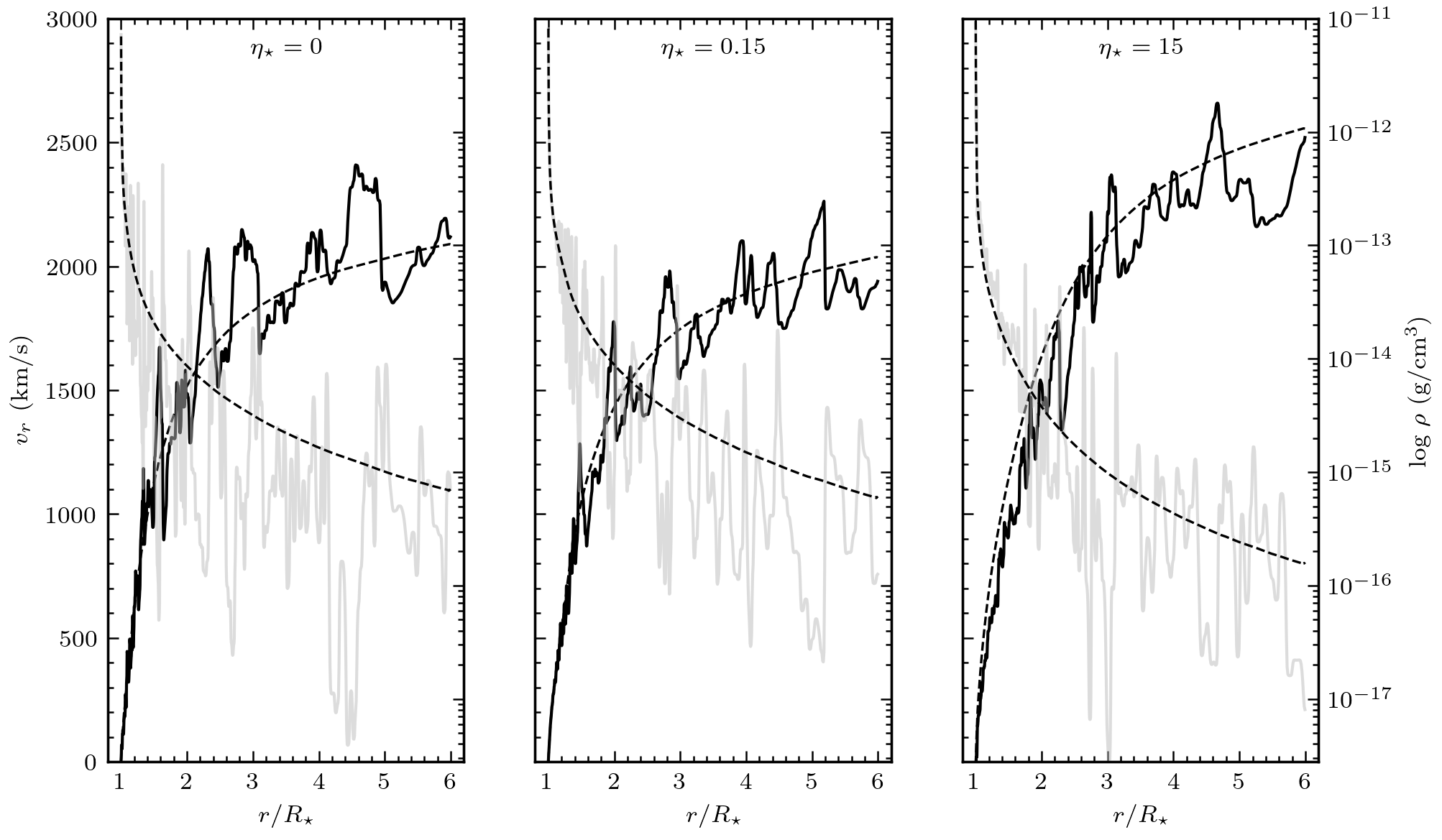}
\caption{Lateral averages of statistically computed wind velocity and wind density (dashed lines). Lateral averages are taken in a wedge near the pole ($\theta=0-45^\circ$, to not have magnetosphere contamination) at simulation termination. Superimposed is a radial cut of wind velocity and density at an arbitrarily chosen colatitude $\theta$ within this wedge (solid lines).}
\label{fig:stat-radcut}
\end{figure*}

To gain insight in the non-linear evolution of the LDI with increasing magnetic confinement it is of interest to consider the wind density. In Fig.~\ref{fig:rhodyn} we contrast a non-magnetic LDI model with weakly and strongly confined LDI models over several subsequent wind evolutionary times ($t=0,50,150,300$\,ks). 

Starting from a steady-state CAK wind, the non-magnetic LDI model develops a rather coherent wind structure initially ($t < 50$\,ks). However, the outwardly accelerating wind structure starts to become progressively disrupted due to Rayleigh--Taylor and thin-shell instabilities \citep{2003A&A...406L...1D} leading to the formation of spatially separated wind clumps. It is worth noting that in this non-magnetic wind model the lateral scale up to which LDI structure forms remains unclear. The fragmentation might result in large lateral scales due to lateral line drag acting on the flow \citep{1990ApJ...349..274R,2020MNRAS.499.4282D} or extend down to smaller lateral scales if, for example, the photosphere consists of many turbulent bubbles that leave their imprint in the wind. Within the one-dimensional SSF this break-up tends to happen up to the lateral Sobolev length such that a near complete lateral incoherence results in the wind. The resulting wind structure shows the typical characteristic slow, overdense clumps separated by a fast, rarefied interclump medium once settled in a `steady-state' \citep{2003A&A...406L...1D,2018A&A...611A..17S}. 

Furthermore since we assume a uniformly bright stellar disc and no photospheric perturbation \citep[e.g.][]{2013MNRAS.428.1837S,2021A&A...648A..79K} all LDI structure seen in our simulations is self-excited. It arises from the back-scattering of photons from the outer wind structure that seeds perturbations closer to the stellar surface which are subsequently amplified by the LDI \citep{2003A&A...406L...1D}.

When introducing a weak magnetic field (i.e.~weak confinement, $\eta_\star=0.15$) the overall picture is not significantly changed. Once sufficiently developed, however, the wind stretches the magnetic field into a nearly radial configuration that is reminiscent of a `split-monopole'. Along with this topology the formation of a current sheet occurs in the magnetic equatorial plane with a corresponding modest density enhancement. The morphology of the wind clumps remains also similar to the non-magnetic case showing that a weak magnetic field has a small effect on the clump dynamics.

The strongly confined LDI wind ($\eta_\star=15$) undergoes a markedly different dynamical evolution compared to the non-magnetic and weakly confined LDI wind. An important difference with respect to the other displayed models is the lack of lateral fragmentation and incoherence of the wind contrary to the standard non-magnetic LDI wind models hitherto performed \citep{2003A&A...406L...1D,2005A&A...437..657D,2018A&A...611A..17S}. This happens particularly in the open field regions where the LDI generated structures manifest themselves as large-scale, wavy shellular sheets that advect outward at the local wind velocity. This suggests that the presence of a strong enough magnetic field makes the LDI unable to fragment into `wind clumps' in this region, but rather transforms it into overdense `wind sheets', still separated by a fast, very rarefied medium. It appears thus that the magnetic field stabilises against lateral plasma motions and forces all LDI structure along the magnetic field line (`frozen-in flow'). Although the physical reasons for the underlying gas motions are different, this situation is quite similar to plasma flow in a sunspot where a strong vertical magnetic field inside the sunspot prohibits convective blobs from fragmenting into the lateral directions.

Additionally, after a long enough time, part of the wind plasma is guided along the magnetic field lines and confined over a small range in latitude by closed magnetic loops. The resulting circumstellar magnetosphere is akin to what has been found in previous magnetic wind studies using the CAK line force \citep[][see also Appendix \ref{sec:comparison}]{2002ApJ...576..413U}. Indeed, plasma at higher latitudes is channelled towards the magnetic equator where collisions from opposite magnetic footprints happen and the resulting plasma compression leads to a rather slow, dense outflow. Most notably there are no signatures of the characteristic LDI structures within the circumstellar magnetosphere, although the wind is radiatively unstable in this region. The apparent absence of LDI-like structures in the magnetosphere might be due to a misalignment between local flow and magnetic field vectors that induces a growth restriction or the fact that higher density material falls back onto the star and dominates the LDI.

The appearance of wind sheets instead of wind clumps also suggests that the wind clumping and porosity properties \citep{2018MNRAS.475..814O} in the open field regions of these magnetic LDI winds might be much different from their non-magnetic counterparts \citep[e.g.][that consider LDI-like `shells' and porosity in context of effects upon absorption of X-rays]{2003A&A...403..217F,2012MNRAS.420.1553S}. What the exact effect of these large-scale shellular wind sheets is on observations remains to be investigated in detail. However, we discuss some first aspects of their presence in Sect.~\ref{sec:obs}.

\subsection{Statistical properties}\label{sec:stats}

To further analyse the LDI structures in our models we consider statistical averages (here temporal averages) of hydrodynamic quantities. These statistics are computed from every numerical iteration beginning significantly after the simulation has developed its characteristic wind structure as set by the dynamical timescale $\tau_\mathrm{dyn}=R_\star/\varv_\infty\approx10$\,ks. All computations start at $t_\mathrm{stat}=200\,\text{ks} \approx 20\tau_\mathrm{dyn}$ for each model such that statistical quantities span $\approx 100\,\text{ks} \approx 10\tau_\mathrm{dyn}$ within the simulations.

To emphasise similarities between our two-dimensional simulations and previous one-dimensional simulations (all treating the LDI line force in a one-dimensional fashion) we display in Fig.~\ref{fig:stat-radcut} a radial cut of wind velocity and density taken at an arbitrarily chosen colatitude $\theta$ away from the magnetosphere. In such an isolated colatitudinal cone the wind properties show the characteristic one-dimensional features of slow, overdense clumps (or shells) that are separated by a fast, nearly-void medium. However, with increasing magnetic confinement the radial velocity and density variations appear less strong. Since all parameters in the models are fixed, except for the polar magnetic field strength, this suggests that the increasing magnetic field strength reduces the strength of the LDI in a relative sense, and as can be seen, also the position at which it sets in. 

Although the discussion so far applies to a single radial wind slice, these typical LDI features occur over a large portion of the wind. To illustrate this we take averages over latitude to recover a corresponding one-dimensional smooth velocity and density profile. In order to have a meaningful average in latitude the simulation domain is constrained to a $45^\circ$ wedge near the pole to avoid material inside the spatial extent of the magnetosphere in the $\eta_\star=15$ case. For direct comparison this wedge is then adopted for each model discussed here. As displayed in Fig.~\ref{fig:stat-radcut} the corresponding smooth mean one-dimensional wind velocity and density describes well the variations in a randomly chosen radial cut. Hence, all radial cuts within the wedge display similar velocity and density variations around this mean and as such are prototypes of a corresponding one-dimensional LDI wind. 

We also point out that under strong confinements the wind becomes faster. This effect is due to the faster-than-radial expansion of the field that results in higher velocities near the magnetic pole \citep{2004ApJ...600.1004O}. Since our wedge is defined near the pole region to leave out any magnetosphere contributions this effect manifests itself inherently in Fig.~\ref{fig:stat-radcut}.

\begin{figure}[t]
\centering
\includegraphics[width=\hsize]{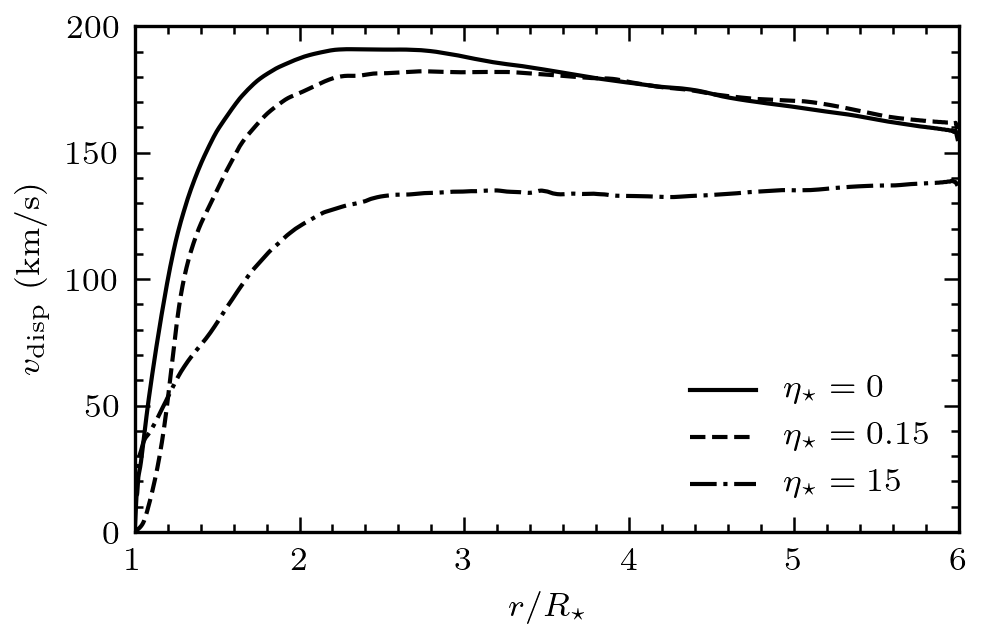}
\caption{Radial velocity dispersion laterally averaged over the wedge near the pole ($\theta=0-45^\circ$).}
\label{fig:vdisp}
\end{figure}

The reduction in LDI strength for increasing magnetic confinement is also seen when considering the velocity dispersion (Fig.~\ref{fig:vdisp}) averaged over the wedge
\begin{equation}
\varv_\mathrm{disp} = \sqrt{\langle \langle  \varv^2 \rangle_t \rangle_\theta - \langle \langle \varv \rangle_t^2 \rangle_\theta },
\end{equation}
which is a good proxy for the reverse shock speed due to the LDI. Since the velocity dispersion relates to the velocity jumps due to shocks this shows that with increasing magnetic confinement the shocks become less strong, resulting in weaker wind density compressions, and to less overdense wind clumps (c.f.~Fig.~\ref{fig:stat-radcut}).

\begin{figure*}[t]
\centering
\includegraphics[width=\hsize]{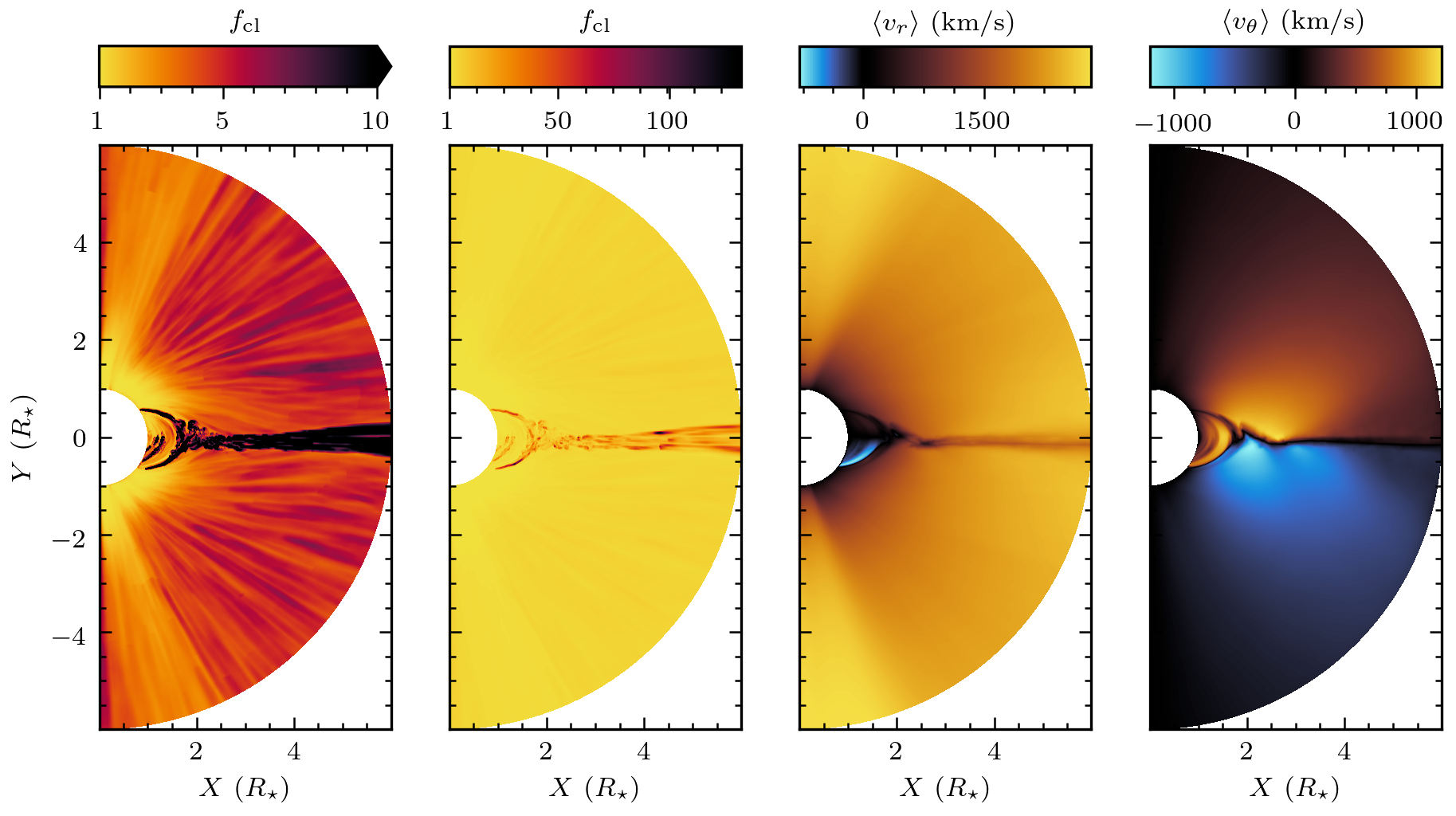}
\caption{Statistical contours (from left to right) of wind clumping restricted to a maximum of ten, full wind clumping, radial wind velocity, and poloidal wind velocity for a magnetically confined LDI wind with $\eta_\star=15$.}
\label{fig:stat-2detas15}
\end{figure*}

In Fig.~\ref{fig:stat-2detas15} we elaborate further on the $\eta_\star=15$ model and show first the wind clumping factor
\begin{equation}
f_\mathrm{cl} = \frac{\langle \rho^2 \rangle_t}{\langle \rho \rangle_t^2}.
\end{equation}
In non-magnetic hot stars wind clumping can seriously affect observational diagnostics depending on $\rho^2$-processes and leads to an overestimate of inferred mass-loss rates with a factor $\sqrt{f_\mathrm{cl}}$. Corrections for $f_\mathrm{cl}$ in non-magnetic hot stars can be obtained via multi-wavelength diagnostics or complex, time-dependent numerical simulations of the LDI. In particular, the optical diagnostic line H$\alpha$ formed at $r\leq 2R_\star$ is often used as a mass-loss rate tracer for non-magnetic hot star winds and a typical wind clumping factor is $f_\mathrm{cl}\approx 10-20$ \citep{2006A&A...454..625P,2011A&A...535A..32N,2021arXiv210808340H,2021arXiv210811734R}. 

For magnetic hot stars, constraints on $f_\mathrm{cl}$ are presently not well established. To that end in Fig.~\ref{fig:stat-2detas15} we display the wind clumping cut off to a maximum of $f_\mathrm{cl}=10$ (typical non-magnetic adopted H$\alpha$ clumping factor) and wind clumping without cut off. Within the magnetosphere density enhancements provide regions with wind clumping $f_\mathrm{cl} \geq 10$ due to fall-back of dense material from the magnetic equatorial plane. Outside the magnetosphere any wind clumping is due to the LDI which is comparatively lower in large portions of the unconfined wind.

Considering wind clumping in the magnetosphere is of interest for wind line-diagnostics such as H$\alpha$ (more discussion in Sect.~\ref{sec:obs}). The wind clumping displayed in Fig.~\ref{fig:stat-2detas15} is calculated from a temporal average, and when further averaging this over the spatial extent of the magnetosphere, we find a modest wind clumping of $f_\mathrm{cl} \approx 4$ inside the magnetosphere. It is likely that such temporal averaging underestimates density enhancements over long times (several hundred thousand iterations in our simulation). In a two-dimensional model there is only one azimuthal slice over which the average is continuously taken leading to a smooth density structure. In three-dimensional simulations, however, at any instant density structures will be arbitrarily distributed along the azimuth in time \citep{2013MNRAS.428.2723U,2019MNRAS.489.3251D}. Therefore, we compute from fifty arbitrarily chosen snapshots a spatial average of wind clumping. Doing so results in magnetospheric wind clumping values of $f_\mathrm{cl} \approx 40$ that is in good agreement with earlier attempts from magnetic line-driven wind models neglecting the LDI \citep{2016MNRAS.462.3830O,2019IAUS..346...45D}.

Alongside $f_\mathrm{cl}$ we also display in Fig.~\ref{fig:stat-2detas15} the average radial and poloidal wind velocity. The radial velocity contours demonstrate that outside the magnetosphere the wind accelerates and averages out to a smooth wind with $\varv_r \approx 2500$\,km\,s$^{-1}$ with a faster flow near the pole \citep{2004ApJ...600.1004O}. Within the magnetosphere on average most material falls back onto the star while only a fraction of it escapes in the magnetic equatorial plane. Similarly, the poloidal velocities show that inside the magnetosphere material from opposite footprints is channelled towards the equatorial plane with $\varv_\theta \approx 700-800$\,km\,s$^{-1}$. Outside the magnetosphere this poloidal velocity gradually decreases towards the pole as magnetic field lines become progressively radially stretched. These velocity contours reinforce that simply assuming a monotonic velocity profile in line-diagnostic studies leads to erroneous wind parameter estimates \citep[e.g.][]{2007MNRAS.381..433H,2015A&A...575A..34M}.

\subsection{Global mass-loss rate}\label{sec:mdotcomp}

For stars with magnetospheres the concept of `mass-loss rate' requires attention because the magnetic channelling makes a significant fraction of material unable to escape the star. Therefore, the mass launched from the stellar surface is really a `mass-feeding rate' $\dot{M}_{B=0}$, which, here, is assumed to be the same mass loss a non-magnetic star undergoes. This quantity, however, can differ significantly from the actual, global mass-loss rate of the star $\dot{M}_{B}$. To first order both mass-loss rate quantities can be related to each other \citep{2008MNRAS.385...97U} by taking into account the area the magnetosphere covers such that the fraction of open magnetic field lines amounts to
\begin{equation}\label{eq:fracmag}
f_B = 1-\sqrt{1-\frac{R_\star}{R_c}}, \qquad R_c \approx R_\star + 0.7(R_A - R_\star)
\end{equation}
and the global mass-loss rate $\dot{M}_B$ scales as
\begin{equation}\label{eq:mdotanalytical}
\dot{M}_B \approx f_B \dot{M}_{B=0}.
\end{equation}

To verify this result, we compute global mass-loss rates from our simulations by considering how much mass is lost through the outer radial boundary. This means we compute in our simulation a mass flux $\langle \rho \varv \rangle_t$, weighted by $\sin \theta$, to average out the variability from the advection of the large-scale wind sheets. 

\begin{figure}[t]
\centering
\includegraphics[width=\hsize]{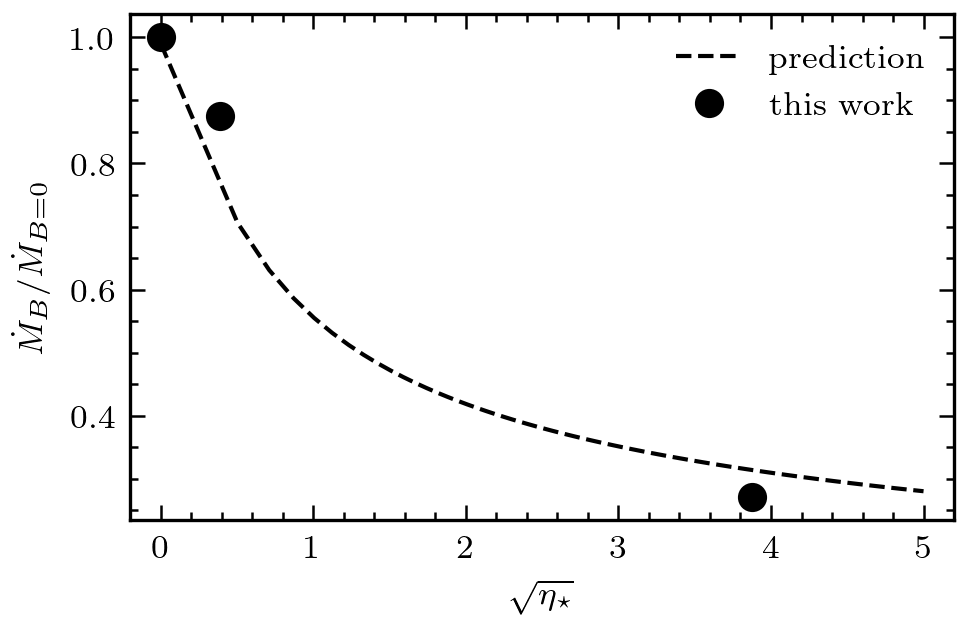}
\caption{Modulation of the global mass-loss rate $\dot{M}_B$ with wind magnetic confinement $\sqrt{\eta_\star} \propto B_p$ for each simulated LDI wind model (solid circles) with the prediction of \citet{2008MNRAS.385...97U} (dashed line).}
\label{fig:mdotcomp}
\end{figure}

Figure \ref{fig:mdotcomp} collects the computed global mass-loss rates and compares them with Eq.~\eqref{eq:mdotanalytical}. Overall there is good agreement between the prediction of \citet{2008MNRAS.385...97U} and our numerical LDI simulations. Such an outcome demonstrates that the LDI force description, when averaged over long enough advection times, resembles well the `steady-state' wind structures also appearing in magnetic CAK models to which \citet{2008MNRAS.385...97U} calibrated Eq.~\eqref{eq:mdotanalytical}.

\subsection{Spatial coherence of wind clumps}\label{sec:spatcoh}

\begin{figure}[t]
\centering
\includegraphics[width=\hsize]{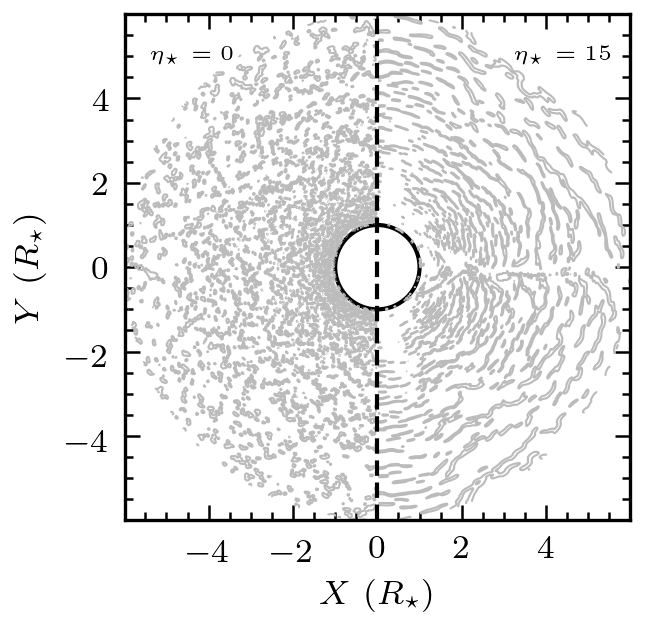}
\caption{Density contrast with respect to the average wind density for the non-magnetic and strongly confined LDI wind model. }
\label{fig:morph}
\end{figure}

To better illustrate the wind morphology in Fig.~\ref{fig:morph} we display a similar wind density state but now normalised as $\rho/\langle \rho \rangle_t = 2$ for the non-magnetic and strongly confined LDI wind. Since the attained density contrasts can become quite large in a shocked LDI wind, the panels in Fig.~\ref{fig:rhodyn} tend to underestimate the amount of structure in the wind (since they are shown at the same density scale). To that end a normalisation with respect to the mean wind density of each LDI model is more appropriate because wind variations occur with respect to this mean. Within this normalisation then the extended shellular-like nature of the wind clumps in the strongly confined LDI wind becomes more apparent showing that these wind sheets (`pancakes') can cover half a hemisphere of the star. On the contrary, the non-magnetic LDI wind still shows small-scale spatially separated wind clumps that tend to elongate at the outer radial boundary due to the spherical divergence.

\section{Discussion on observational signatures and stellar evolution}\label{sec:obs}

We discuss here some possibilities to disentangle the signatures of the wind sheets as found within the numerical simulations. We also comment on any effects of the predicted mass-loss rates and stellar evolution modelling of magnetic hot stars. The discussion focuses on the $\eta_\star=15$ LDI model since this is a reasonably realistic wind confinement for magnetic O-stars in the Galaxy.

\subsection{Spectroscopy}

Over the past decade it has been extensively demonstrated that the circumstellar magnetosphere is an excellent probe to infer several wind and magnetospheric parameters of magnetic hot stars using X-rays \citep{2010A&A...520A..59N,2012MNRAS.423.3413N,2016ApJ...831..138N}, resonance UV lines \citep{2013MNRAS.431.2253M,2015MNRAS.452.2641N,2019MNRAS.483.2814D,2021MNRAS.506.5373E}, and optical recombination lines \citep{2012MNRAS.423L..21S,2012MNRAS.419.2459W,2013MNRAS.428.2723U,2015MNRAS.447.2551W,2016MNRAS.462.3830O,2020MNRAS.499.5379S}. These prime diagnostic domains can be complemented with the infrared \citep{2014ApJ...784L..30E,2015A&A...578A.112O} and radio domain \citep{2004A&A...418..593T,2020ApJ...900..156D}. 

The LDI generates strong reverse shocks that leads to X-ray emission in line-driven winds. In particular, magnetic O-stars are strong sources of soft X-ray emission \citep{2014ApJS..215...10N} that can naturally result from the wind sheets outside the magnetosphere. Nonetheless, it is worth noting that so far in all numerical models LDI generated shocks underestimate the amount of soft X-ray emission in non-magnetic line-driven winds \citep{1997A&A...322..878F}. Therefore, within the current modelling attempts, it is questionable whether the magnetic LDI models can correctly estimate the amount of soft X-ray emission for magnetic hot stars. On the other hand, it seems unlikely that the LDI contributes significantly to the generation of hard X-rays that are understood to emerge from shocks resulting from the collision of magnetically channelled flow \citep{1997A&A...323..121B,2005ApJ...628..986G}. 

The UV line column-mass dependency could provide insights into the advective motion of the wind sheets. This could appear quite similar, for example, to the discrete absorption components (DACs) that are seen in absorption troughs of UV P-Cygni profiles of (magnetic) hot stars \citep{1989ApJS...69..527H,1996A&AS..116..257K,2015A&A...574A.142M,2016A&A...594A..56S,2017MNRAS.470.3672D}. However, it is important to point out that such DACs are believed to be modulated by stellar rotation, whereas here we would expect variation on a dynamical advection timescale, which thus should be different. For stars viewed near their magnetic pole, the advective motion of the wind sheets can manifest itself in the absorption trough of such P-Cygni profiles. Since the sheets typically traverse one stellar radius in a few hours this does require high-cadence and high-resolution spectroscopy assuming observational noise does not hide the signature. Further insights into this hypothesis, however, can be obtained using synthetic observations of UV resonance lines using state-of-the-art radiative transfer codes \citep[e.g.][]{2020A&A...633A..16H}.

Finally, the H$\alpha$ line-profile variability in magnetic hot stars stems naturally from the density structure within the magnetosphere \citep[e.g.][]{2012MNRAS.423L..21S}. Since H$\alpha$ is a recombination line the inference of mass-feeding rates crucially depends on the amount of wind clumping taken into account. For example, in non-magnetic hot stars the commonly adopted value is $f_\mathrm{cl} = 10$, but as seen in Fig.~\ref{fig:stat-2detas15} such wind clumping factors can be much higher in the magnetically confined region. A typical spatial average of wind clumping inside the magnetosphere amounts to $f_\mathrm{cl} \approx 40$ (but see discussion in Sect.~\ref{sec:stats}). This vividly illustrates that applying typical wind clumping factors used for non-magnetic hot stars may provide erroneous estimates of the amount and effect of wind clumping. Consequently the mass-feeding rate of such magnetic hot stars may also be more reduced when compared with a non-magnetic counterpart. This reduction in mass-feeding rate can also have important effects on the stellar evolution and final fate of magnetic hot stars (see below).

\subsection{Photometry and polarimetry}

Current photometric data has shown to be another alternative to diagnose hot star magnetospheres \citep{2019MNRAS.487..304D,2020MNRAS.492.1199M}. Such photometric signatures might be harder to relate to an inhomogeneous wind, however, since other processes can disturb the signal (e.g.~stellar pulsations or rotational modulations). For example, the wind sheet overdensities near the star can potentially be seen as semi-regular photometric variability. We expect this effect to be most pronounced near the magnetic pole of the star where the wind flow is mainly radial and does not suffer from magnetospheric contamination. Any possible photometric signatures would then quasi-periodically appear in the photometric data within a time span of several hours needed to advect outwards. 

Due to the high degree of ionisation in line-driven winds measurements of continuum polarisation can provide another window for probing the existence of wind sheets. Indeed, the linearly polarised light due to electron scattering in these hot atmospheres can provide a clue into the geometrical distribution of the wind sheets, that is deviations from spherical symmetry.

\subsection{Evolutionary modelling of magnetic hot stars}

Determination of the mass-feeding rates versus the mass-loss rates of magnetic hot stars is of interest due to their potentially different evolutionary pathways compared to non-magnetic hot stars. Particularly, it has been shown that magnetic OB stars quench a large amount of their mass flux and could become progenitors of heavy stellar-mass black holes nowadays linked with gravitational wave detections \citep{2017MNRAS.466.1052P}. 

In these stellar evolution models of magnetic hot stars the assumed mass-loss rate (i.e.~the mass-feeding rate $\dot{M}_{B=0}$) has been taken from corresponding non-magnetic OB star predictions \citep{2001A&A...369..574V}. This $\dot{M}_{B=0}$ is then linked via Eq.~\eqref{eq:mdotanalytical} to the global mass-loss rate $\dot{M}_B$ in such evolution models. However, it is worth noting that the recent line-driven wind models of \citet{2019A&A...632A.126S} and \citet{2021A&A...648A..36B} predict a factor three lower $\dot{M}_{B=0}$ than \citet{2001A&A...369..574V} in the O-star regime. Therefore, using these new O-star mass loss rate predictions would also reduce $\dot{M}_{B}$ by a factor of three compared to what is currently employed in magnetic OB star evolution modelling \citep{2017MNRAS.466.1052P,2017A&A...599L...5G,2019MNRAS.485.5843K,2020MNRAS.493..518K}. Given the strong interdependence of mass loss rate and evolution for hot stars \citep{2000A&A...361..101M}, further studies are required to test the effects of this further reduction in mass-loss for magnetic OB stars.

It is important to recall that the inclusion of the LDI does not alter either the global mass-feeding rate as predicted from line-driven wind theory, nor the mass-loss rate predicted from magnetospheric confinement (see discussion surrounding Fig.~\ref{fig:mdotcomp}). However, for observational modelling, constraints on wind clumping (inside and outside of the magnetosphere) are necessary to further understand and constrain both the mass-feeding and mass-loss rate of the star. For example, as discussed above, the validity of using non-magnetic mass-feeding rates in stellar evolution modelling is a priori not guaranteed due to the additional (unknown) complexities the magnetic field can exert on the mass flux. However, first empirical constraints from H$\alpha$ line-profile modelling within the circumstellar magnetosphere, taking into account wind clumping corrections a posteriori, provide a tentative back-up for these non-magnetic mass-feeding rate adaptations \citep{2019IAUS..346...45D}. In particular, these authors demonstrated that for a selected sample of Galactic magnetic OB stars the adaptation of the \citet{2001A&A...369..574V} mass-loss rates is justifiable if the global mass-loss rate $\dot{M}_B$ is scaled down accordingly with a factor of $f_\mathrm{cl} \approx 50$, which in light of the present study, aligns well with the results discussed in Sect.~\ref{sec:stats}. Future extensions to this study in light of the wind clumping outside the magnetosphere found here could, however, test this finding with additional diagnostics.

\section{Conclusions and future work}\label{sec:conc}

In this paper we have presented a first theoretical investigation into the wind `clumping' properties of magnetic O-stars due to the line-deshadowing instability (LDI). In particular we have employed two-dimensional magnetohydrodynamic simulations whereby the radiation line force is still computed using the one-dimensional Smooth Source Function (SSF) approximation \citep{1996ApJ...462..894O}.

Our main result is that increasing the wind-magnetic confinement to values typically expected for magnetic O-stars, that is magnetically confined flows, leads to a two-state wind morphology; (i) in accordance with previous magnetic CAK wind models inside the circumstellar magnetosphere the flow is confined with regular wind flow fall-back towards the star, and (ii) outside the magnetosphere large-scale coherent, shellular `wind sheets' separated by a rarefied medium advect outwards contrary to previous magnetic CAK and non-magnetic LDI wind models. 

We are currently extending our radiation-MHD models to take into account a proper (although spatially restricted) description of the two-dimensional SSF radiation transport \citep{2018A&A...611A..17S}. This allows us to further study the influence of strong magnetic fields and the fragmentation of LDI-like structures with the inclusion of non-radial line forces.

Among several proposed potential observational signatures, we are also investigating the effects of UV line-profile variability from LDI generated wind sheets using a state-of-the-art three-dimensional radiative transfer code \citep{2020A&A...633A..16H}. Since UV lines are column-mass diagnostics the advection of overdense wind sheets near the pole might be seen within the absorption trough of a P-Cygni profile. Whether such effect are observable, that is not dominated by noise, remains to be investigated. Results of these works will be reported in the future.

Overall the main conclusion of the present study is that LDI generated wind structure in a magnetically confined line-driven wind can differ significantly with non-magnetic LDI wind models, as well as with magnetic line-driven wind models that apply CAK theory for the description of the radiation line force (i.e.~the intrinsic appearance of wind sheets due to the LDI not seen in the other models). With future research we intend to further investigate theoretical and observational aspects to test the signatures predicted from the present simulations and assess its predictive power and validity.

\begin{acknowledgements}
It is a pleasure to thank Stan Owocki and Asif ud-Doula for interesting discussions. FAD and JOS acknowledge support from the Odysseus program of the Belgian Research Foundation Flanders (FWO) under grant G0H9218N. NDK acknowledges support from the KU Leuven C1 grant MAESTRO C16/17/007. The resources and services used in parts of this work were provided by the VSC (Flemish Supercomputer Center), funded by the Research Foundation - Flanders (FWO) and the Flemish Government. The \textsc{cmasher} library has been used to provide perceptually uniform, colour-blind friendly colour maps in this work \citep{2020JOSS....5.2004V}. We thank the referee, Prof. Achim Feldmeier, for constructive criticism and comments.
\end{acknowledgements}

%
%

\bibliography{refs} 

\begin{thebibliography}{112}
\expandafter\ifx\csname natexlab\endcsname\relax\def\natexlab#1{#1}\fi

\bibitem[{{Alecian} {et~al.}(2015){Alecian}, {Neiner}, {Wade}, {Mathis},
  {Bohlender}, {C{\'e}bron}, {Folsom}, {Grunhut}, {Le Bouquin}, {Petit},
  {Sana}, {Tkachenko}, \& {ud-Doula}}]{2015IAUS..307..330A}
{Alecian}, E., {Neiner}, C., {Wade}, G.~A., {et~al.} 2015, in New Windows on
  Massive Stars, ed. G.~{Meynet}, C.~{Georgy}, J.~{Groh}, \& P.~{Stee}, Vol.
  307, 330--335

\bibitem[{{Alecian} {et~al.}(2017){Alecian}, {Villebrun}, {Grunhut}, {Hussain},
  {Neiner}, \& {Wade}}]{2017arXiv170510650A}
{Alecian}, E., {Villebrun}, F., {Grunhut}, J., {et~al.} 2017, arXiv e-prints,
  arXiv:1705.10650

\bibitem[{{Babel} \& {Montmerle}(1997)}]{1997A&A...323..121B}
{Babel}, J. \& {Montmerle}, T. 1997, \aap, 323, 121

\bibitem[{{Bagnulo} {et~al.}(2020){Bagnulo}, {Wade}, {Naz{\'e}}, {Grunhut},
  {Shultz}, {Asher}, {Crowther}, {Evans}, {David-Uraz}, {Howarth}, {Morrell},
  {Munoz}, {Neiner}, {Puls}, {Szyma{\'n}ski}, \& {Vink}}]{2020A&A...635A.163B}
{Bagnulo}, S., {Wade}, G.~A., {Naz{\'e}}, Y., {et~al.} 2020, \aap, 635, A163

\bibitem[{{Bard} \& {Townsend}(2016)}]{2016MNRAS.462.3672B}
{Bard}, C. \& {Townsend}, R. H.~D. 2016, \mnras, 462, 3672

\bibitem[{{Bj{\"o}rklund} {et~al.}(2021){Bj{\"o}rklund}, {Sundqvist}, {Puls},
  \& {Najarro}}]{2021A&A...648A..36B}
{Bj{\"o}rklund}, R., {Sundqvist}, J.~O., {Puls}, J., \& {Najarro}, F. 2021,
  \aap, 648, A36

\bibitem[{{Blondin} {et~al.}(1990){Blondin}, {Kallman}, {Fryxell}, \&
  {Taam}}]{1990ApJ...356..591B}
{Blondin}, J.~M., {Kallman}, T.~R., {Fryxell}, B.~A., \& {Taam}, R.~E. 1990,
  \apj, 356, 591

\bibitem[{{Borra} {et~al.}(1982){Borra}, {Landstreet}, \&
  {Mestel}}]{1982ARA&A..20..191B}
{Borra}, E.~F., {Landstreet}, J.~D., \& {Mestel}, L. 1982, \araa, 20, 191

\bibitem[{{Carlberg}(1980)}]{1980ApJ...241.1131C}
{Carlberg}, R.~G. 1980, \apj, 241, 1131

\bibitem[{{Castor} {et~al.}(1975){Castor}, {Abbott}, \&
  {Klein}}]{1975ApJ...195..157C}
{Castor}, J.~I., {Abbott}, D.~C., \& {Klein}, R.~I. 1975, \apj, 195, 157

\bibitem[{{Colella} \& {Woodward}(1984)}]{1984JCoPh..54..174C}
{Colella}, P. \& {Woodward}, P.~R. 1984, Journal of Computational Physics, 54,
  174

\bibitem[{{Cranmer} \& {Owocki}(1995)}]{1995ApJ...440..308C}
{Cranmer}, S.~R. \& {Owocki}, S.~P. 1995, \apj, 440, 308

\bibitem[{{Daley-Yates} {et~al.}(2019){Daley-Yates}, {Stevens}, \&
  {ud-Doula}}]{2019MNRAS.489.3251D}
{Daley-Yates}, S., {Stevens}, I.~R., \& {ud-Doula}, A. 2019, \mnras, 489, 3251

\bibitem[{{Das} {et~al.}(2020){Das}, {Mondal}, \&
  {Chandra}}]{2020ApJ...900..156D}
{Das}, B., {Mondal}, S., \& {Chandra}, P. 2020, \apj, 900, 156

\bibitem[{{David-Uraz} {et~al.}(2019{\natexlab{a}}){David-Uraz}, {Erba},
  {Petit}, {Fullerton}, {Martins}, {Walborn}, {MacInnis}, {Barb{\'a}}, {Cohen},
  {Ma{\'\i}z Apell{\'a}niz}, {Naz{\'e}}, {Owocki}, {Sundqvist}, {ud-Doula}, \&
  {Wade}}]{2019MNRAS.483.2814D}
{David-Uraz}, A., {Erba}, C., {Petit}, V., {et~al.} 2019{\natexlab{a}}, \mnras,
  483, 2814

\bibitem[{{David-Uraz} {et~al.}(2019{\natexlab{b}}){David-Uraz}, {Neiner},
  {Sikora}, {Bowman}, {Petit}, {Chowdhury}, {Handler}, {Pergeorelis},
  {Cantiello}, {Cohen}, {Erba}, {Keszthelyi}, {Khalack}, {Kobzar}, {Kochukhov},
  {Labadie-Bartz}, {Lovekin}, {MacInnis}, {Owocki}, {Pablo}, {Shultz},
  {ud-Doula}, \& {Wade}}]{2019MNRAS.487..304D}
{David-Uraz}, A., {Neiner}, C., {Sikora}, J., {et~al.} 2019{\natexlab{b}},
  \mnras, 487, 304

\bibitem[{{David-Uraz} {et~al.}(2017){David-Uraz}, {Owocki}, {Wade},
  {Sundqvist}, \& {Kee}}]{2017MNRAS.470.3672D}
{David-Uraz}, A., {Owocki}, S.~P., {Wade}, G.~A., {Sundqvist}, J.~O., \& {Kee},
  N.~D. 2017, \mnras, 470, 3672

\bibitem[{{Dessart} \& {Owocki}(2003)}]{2003A&A...406L...1D}
{Dessart}, L. \& {Owocki}, S.~P. 2003, \aap, 406, L1

\bibitem[{{Dessart} \& {Owocki}(2005)}]{2005A&A...437..657D}
{Dessart}, L. \& {Owocki}, S.~P. 2005, \aap, 437, 657

\bibitem[{{Driessen} {et~al.}(2020){Driessen}, {Kee}, {Sundqvist}, \&
  {Owocki}}]{2020MNRAS.499.4282D}
{Driessen}, F.~A., {Kee}, N.~D., {Sundqvist}, J.~O., \& {Owocki}, S.~P. 2020,
  \mnras, 499, 4282

\bibitem[{{Driessen} {et~al.}(2019{\natexlab{a}}){Driessen}, {Sundqvist}, \&
  {Kee}}]{2019A&A...631A.172D}
{Driessen}, F.~A., {Sundqvist}, J.~O., \& {Kee}, N.~D. 2019{\natexlab{a}},
  \aap, 631, A172

\bibitem[{{Driessen} {et~al.}(2019{\natexlab{b}}){Driessen}, {Sundqvist}, \&
  {Wade}}]{2019IAUS..346...45D}
{Driessen}, F.~A., {Sundqvist}, J.~O., \& {Wade}, G.~A. 2019{\natexlab{b}}, IAU
  Symposium, 346, 45

\bibitem[{{Dyda} \& {Proga}(2018)}]{2018MNRAS.481.5263D}
{Dyda}, S. \& {Proga}, D. 2018, \mnras, 481, 5263

\bibitem[{{Eikenberry} {et~al.}(2014){Eikenberry}, {Chojnowski}, {Wisniewski},
  {Majewski}, {Shetrone}, {Whelan}, {Bizyaev}, {Borish}, {Davenport}, {Ebelke},
  {Feuillet}, {Frinchaboy}, {Garner}, {Hearty}, {Holtzman}, {Li},
  {M{\'e}sz{\'a}ros}, {Nidever}, {Schneider}, {Skrutskie}, {Wilson}, \&
  {Zasowski}}]{2014ApJ...784L..30E}
{Eikenberry}, S.~S., {Chojnowski}, S.~D., {Wisniewski}, J., {et~al.} 2014,
  \apjl, 784, L30

\bibitem[{{El Mellah} \& {Casse}(2017)}]{2017MNRAS.467.2585E}
{El Mellah}, I. \& {Casse}, F. 2017, \mnras, 467, 2585

\bibitem[{{Erba} {et~al.}(2021){Erba}, {David-Uraz}, {Petit}, {Hennicker},
  {Fletcher}, {Fullerton}, {Naz{\'e}}, {Sundqvist}, \&
  {ud-Doula}}]{2021MNRAS.506.5373E}
{Erba}, C., {David-Uraz}, A., {Petit}, V., {et~al.} 2021, \mnras, 506, 5373

\bibitem[{{Feldmeier}(1995)}]{1995A&A...299..523F}
{Feldmeier}, A. 1995, \aap, 299, 523

\bibitem[{{Feldmeier} {et~al.}(2003){Feldmeier}, {Oskinova}, \&
  {Hamann}}]{2003A&A...403..217F}
{Feldmeier}, A., {Oskinova}, L., \& {Hamann}, W.~R. 2003, \aap, 403, 217

\bibitem[{{Feldmeier} {et~al.}(1997){Feldmeier}, {Puls}, \&
  {Pauldrach}}]{1997A&A...322..878F}
{Feldmeier}, A., {Puls}, J., \& {Pauldrach}, A.~W.~A. 1997, \aap, 322, 878

\bibitem[{{Feldmeier} \& {Thomas}(2017)}]{2017MNRAS.469.3102F}
{Feldmeier}, A. \& {Thomas}, T. 2017, \mnras, 469, 3102

\bibitem[{{Fossati} {et~al.}(2015){Fossati}, {Castro}, {Sch{\"o}ller},
  {Hubrig}, {Langer}, {Morel}, {Briquet}, {Herrero}, {Przybilla}, {Sana},
  {Schneider}, {de Koter}, \& {BOB Collaboration}}]{2015A&A...582A..45F}
{Fossati}, L., {Castro}, N., {Sch{\"o}ller}, M., {et~al.} 2015, \aap, 582, A45

\bibitem[{{Friend} \& {Abbott}(1986)}]{1986ApJ...311..701F}
{Friend}, D.~B. \& {Abbott}, D.~C. 1986, \apj, 311, 701

\bibitem[{{Friend} \& {MacGregor}(1984)}]{1984ApJ...282..591F}
{Friend}, D.~B. \& {MacGregor}, K.~B. 1984, \apj, 282, 591

\bibitem[{{Gagn{\'e}} {et~al.}(2005){Gagn{\'e}}, {Oksala}, {Cohen}, {Tonnesen},
  {ud-Doula}, {Owocki}, {Townsend}, \& {MacFarlane}}]{2005ApJ...628..986G}
{Gagn{\'e}}, M., {Oksala}, M.~E., {Cohen}, D.~H., {et~al.} 2005, \apj, 628, 986

\bibitem[{{Gayley}(1995)}]{1995ApJ...454..410G}
{Gayley}, K.~G. 1995, \apj, 454, 410

\bibitem[{{Georgy} {et~al.}(2017){Georgy}, {Meynet}, {Ekstr{\"o}m}, {Wade},
  {Petit}, {Keszthelyi}, \& {Hirschi}}]{2017A&A...599L...5G}
{Georgy}, C., {Meynet}, G., {Ekstr{\"o}m}, S., {et~al.} 2017, \aap, 599, L5

\bibitem[{{Gottlieb} \& {Shu}(1998)}]{1998MaCom..67...73G}
{Gottlieb}, S. \& {Shu}, C.~W. 1998, Mathematics of Computation, 67, 73

\bibitem[{{Grunhut} {et~al.}(2017){Grunhut}, {Wade}, {Neiner}, {Oksala},
  {Petit}, {Alecian}, {Bohlender}, {Bouret}, {Henrichs}, {Hussain},
  {Kochukhov}, \& {MiMeS Collaboration}}]{2017MNRAS.465.2432G}
{Grunhut}, J.~H., {Wade}, G.~A., {Neiner}, C., {et~al.} 2017, \mnras, 465, 2432

\bibitem[{Harten {et~al.}(1983)Harten, Lax, \& Leer}]{HLL}
Harten, A., Lax, P.~D., \& Leer, B.~v. 1983, SIAM Review, 25, 35

\bibitem[{{Hawcroft} {et~al.}(2021){Hawcroft}, {Sana}, {Mahy}, {Sundqvist},
  {Abdul-Masih}, {Bouret}, {Brands}, {de Koter}, {Driessen}, \&
  {Puls}}]{2021arXiv210808340H}
{Hawcroft}, C., {Sana}, H., {Mahy}, L., {et~al.} 2021, arXiv e-prints,
  arXiv:2108.08340

\bibitem[{Hedstrom(1979)}]{HEDSTROM1979222}
Hedstrom, G. 1979, Journal of Computational Physics, 30, 222

\bibitem[{{Hennicker} {et~al.}(2020){Hennicker}, {Puls}, {Kee}, \&
  {Sundqvist}}]{2020A&A...633A..16H}
{Hennicker}, L., {Puls}, J., {Kee}, N.~D., \& {Sundqvist}, J.~O. 2020, \aap,
  633, A16

\bibitem[{{Howarth} \& {Prinja}(1989)}]{1989ApJS...69..527H}
{Howarth}, I.~D. \& {Prinja}, R.~K. 1989, \apjs, 69, 527

\bibitem[{{Howarth} {et~al.}(2007){Howarth}, {Walborn}, {Lennon}, {Puls},
  {Naz{\'e}}, {Annuk}, {Antokhin}, {Bohlender}, {Bond}, {Donati}, {Georgiev},
  {Gies}, {Harmer}, {Herrero}, {Kolka}, {McDavid}, {Morel}, {Negueruela},
  {Rauw}, \& {Reig}}]{2007MNRAS.381..433H}
{Howarth}, I.~D., {Walborn}, N.~R., {Lennon}, D.~J., {et~al.} 2007, \mnras,
  381, 433

\bibitem[{{Kaper} {et~al.}(1996){Kaper}, {Henrichs}, {Nichols}, {Snoek},
  {Volten}, \& {Zwarthoed}}]{1996A&AS..116..257K}
{Kaper}, L., {Henrichs}, H.~F., {Nichols}, J.~S., {et~al.} 1996, \aaps, 116,
  257

\bibitem[{{Kee} {et~al.}(2016){Kee}, {Owocki}, \&
  {Sundqvist}}]{2016MNRAS.458.2323K}
{Kee}, N.~D., {Owocki}, S., \& {Sundqvist}, J.~O. 2016, \mnras, 458, 2323

\bibitem[{Keppens {et~al.}(2021)Keppens, Teunissen, Xia, \&
  Porth}]{KEPPENS2021316}
Keppens, R., Teunissen, J., Xia, C., \& Porth, O. 2021, Computers \&
  Mathematics with Applications, 81, 316

\bibitem[{{Keszthelyi} {et~al.}(2019){Keszthelyi}, {Meynet}, {Georgy}, {Wade},
  {Petit}, \& {David-Uraz}}]{2019MNRAS.485.5843K}
{Keszthelyi}, Z., {Meynet}, G., {Georgy}, C., {et~al.} 2019, \mnras, 485, 5843

\bibitem[{{Keszthelyi} {et~al.}(2020){Keszthelyi}, {Meynet}, {Shultz},
  {David-Uraz}, {ud-Doula}, {Townsend}, {Wade}, {Georgy}, {Petit}, \&
  {Owocki}}]{2020MNRAS.493..518K}
{Keszthelyi}, Z., {Meynet}, G., {Shultz}, M.~E., {et~al.} 2020, \mnras, 493,
  518

\bibitem[{{Krti{\v{c}}ka} \& {Feldmeier}(2021)}]{2021A&A...648A..79K}
{Krti{\v{c}}ka}, J. \& {Feldmeier}, A. 2021, \aap, 648, A79

\bibitem[{{K{\"u}ker}(2017)}]{2017AN....338..868K}
{K{\"u}ker}, M. 2017, Astronomische Nachrichten, 338, 868

\bibitem[{{Lagae} {et~al.}(2021){Lagae}, {Driessen}, {Hennicker}, {Kee}, \&
  {Sundqvist}}]{2021A&A...648A..94L}
{Lagae}, C., {Driessen}, F.~A., {Hennicker}, L., {Kee}, N.~D., \& {Sundqvist},
  J.~O. 2021, \aap, 648, A94

\bibitem[{{MacGregor} {et~al.}(1979){MacGregor}, {Hartmann}, \&
  {Raymond}}]{1979ApJ...231..514M}
{MacGregor}, K.~B., {Hartmann}, L., \& {Raymond}, J.~C. 1979, \apj, 231, 514

\bibitem[{{Marcolino} {et~al.}(2013){Marcolino}, {Bouret}, {Sundqvist},
  {Walborn}, {Fullerton}, {Howarth}, {Wade}, \&
  {ud-Doula}}]{2013MNRAS.431.2253M}
{Marcolino}, W.~L.~F., {Bouret}, J.~C., {Sundqvist}, J.~O., {et~al.} 2013,
  \mnras, 431, 2253

\bibitem[{{Martins} {et~al.}(2015{\natexlab{a}}){Martins}, {Herv{\'e}},
  {Bouret}, {Marcolino}, {Wade}, {Neiner}, {Alecian}, {Grunhut}, \&
  {Petit}}]{2015A&A...575A..34M}
{Martins}, F., {Herv{\'e}}, A., {Bouret}, J.~C., {et~al.} 2015{\natexlab{a}},
  \aap, 575, A34

\bibitem[{{Martins} {et~al.}(2015{\natexlab{b}}){Martins}, {Marcolino},
  {Hillier}, {Donati}, \& {Bouret}}]{2015A&A...574A.142M}
{Martins}, F., {Marcolino}, W., {Hillier}, D.~J., {Donati}, J.~F., \& {Bouret},
  J.~C. 2015{\natexlab{b}}, \aap, 574, A142

\bibitem[{{Mestel}(1968)}]{1968MNRAS.138..359M}
{Mestel}, L. 1968, \mnras, 138, 359

\bibitem[{{Meynet} \& {Maeder}(2000)}]{2000A&A...361..101M}
{Meynet}, G. \& {Maeder}, A. 2000, \aap, 361, 101

\bibitem[{{Munoz} {et~al.}(2020){Munoz}, {Wade}, {Naz{\'e}}, {Puls}, {Bagnulo},
  \& {Szyma{\'n}ski}}]{2020MNRAS.492.1199M}
{Munoz}, M.~S., {Wade}, G.~A., {Naz{\'e}}, Y., {et~al.} 2020, \mnras, 492, 1199

\bibitem[{{Najarro} {et~al.}(2011){Najarro}, {Hanson}, \&
  {Puls}}]{2011A&A...535A..32N}
{Najarro}, F., {Hanson}, M.~M., \& {Puls}, J. 2011, \aap, 535, A32

\bibitem[{{Naz{\'e}} {et~al.}(2012){Naz{\'e}}, {Bagnulo}, {Petit}, {Rivinius},
  {Wade}, {Rauw}, \& {Gagn{\'e}}}]{2012MNRAS.423.3413N}
{Naz{\'e}}, Y., {Bagnulo}, S., {Petit}, V., {et~al.} 2012, \mnras, 423, 3413

\bibitem[{{Naz{\'e}} {et~al.}(2014){Naz{\'e}}, {Petit}, {Rinbrand}, {Cohen},
  {Owocki}, {ud-Doula}, \& {Wade}}]{2014ApJS..215...10N}
{Naz{\'e}}, Y., {Petit}, V., {Rinbrand}, M., {et~al.} 2014, \apjs, 215, 10

\bibitem[{{Naz{\'e}} {et~al.}(2015){Naz{\'e}}, {Sundqvist}, {Fullerton},
  {ud-Doula}, {Wade}, {Rauw}, \& {Walborn}}]{2015MNRAS.452.2641N}
{Naz{\'e}}, Y., {Sundqvist}, J.~O., {Fullerton}, A.~W., {et~al.} 2015, \mnras,
  452, 2641

\bibitem[{{Naz{\'e}} {et~al.}(2010){Naz{\'e}}, {Ud-Doula}, {Spano}, {Rauw}, {De
  Becker}, \& {Walborn}}]{2010A&A...520A..59N}
{Naz{\'e}}, Y., {Ud-Doula}, A., {Spano}, M., {et~al.} 2010, \aap, 520, A59

\bibitem[{{Naz{\'e}} {et~al.}(2016){Naz{\'e}}, {ud-Doula}, \&
  {Zhekov}}]{2016ApJ...831..138N}
{Naz{\'e}}, Y., {ud-Doula}, A., \& {Zhekov}, S.~A. 2016, \apj, 831, 138

\bibitem[{{Oksala} {et~al.}(2015){Oksala}, {Grunhut}, {Kraus}, {Borges
  Fernandes}, {Neiner}, {Condori}, {Campagnolo}, \&
  {Souza}}]{2015A&A...578A.112O}
{Oksala}, M.~E., {Grunhut}, J.~H., {Kraus}, M., {et~al.} 2015, \aap, 578, A112

\bibitem[{{Owocki} {et~al.}(1988){Owocki}, {Castor}, \&
  {Rybicki}}]{1988ApJ...335..914O}
{Owocki}, S.~P., {Castor}, J.~I., \& {Rybicki}, G.~B. 1988, \apj, 335, 914

\bibitem[{{Owocki} \& {Puls}(1996)}]{1996ApJ...462..894O}
{Owocki}, S.~P. \& {Puls}, J. 1996, \apj, 462, 894

\bibitem[{{Owocki} \& {Rybicki}(1984)}]{1984ApJ...284..337O}
{Owocki}, S.~P. \& {Rybicki}, G.~B. 1984, \apj, 284, 337

\bibitem[{{Owocki} \& {Sundqvist}(2018)}]{2018MNRAS.475..814O}
{Owocki}, S.~P. \& {Sundqvist}, J.~O. 2018, \mnras, 475, 814

\bibitem[{{Owocki} \& {ud-Doula}(2004)}]{2004ApJ...600.1004O}
{Owocki}, S.~P. \& {ud-Doula}, A. 2004, \apj, 600, 1004

\bibitem[{{Owocki} {et~al.}(2016){Owocki}, {ud-Doula}, {Sundqvist}, {Petit},
  {Cohen}, \& {Townsend}}]{2016MNRAS.462.3830O}
{Owocki}, S.~P., {ud-Doula}, A., {Sundqvist}, J.~O., {et~al.} 2016, \mnras,
  462, 3830

\bibitem[{{Pauldrach} {et~al.}(1986){Pauldrach}, {Puls}, \&
  {Kudritzki}}]{1986A&A...164...86P}
{Pauldrach}, A., {Puls}, J., \& {Kudritzki}, R.~P. 1986, \aap, 164, 86

\bibitem[{{Petit} {et~al.}(2017){Petit}, {Keszthelyi}, {MacInnis}, {Cohen},
  {Townsend}, {Wade}, {Thomas}, {Owocki}, {Puls}, \&
  {ud-Doula}}]{2017MNRAS.466.1052P}
{Petit}, V., {Keszthelyi}, Z., {MacInnis}, R., {et~al.} 2017, \mnras, 466, 1052

\bibitem[{{Petit} {et~al.}(2013){Petit}, {Owocki}, {Wade}, {Cohen},
  {Sundqvist}, {Gagn{\'e}}, {Ma{\'{\i}}z Apell{\'a}niz}, {Oksala}, {Bohlender},
  {Rivinius}, {Henrichs}, {Alecian}, {Townsend}, {ud-Doula}, \& {MiMeS
  Collaboration}}]{2013MNRAS.429..398P}
{Petit}, V., {Owocki}, S.~P., {Wade}, G.~A., {et~al.} 2013, \mnras, 429, 398

\bibitem[{{Petit} {et~al.}(2019){Petit}, {Wade}, {Schneider}, {Fossati},
  {Kamp}, {Neiner}, {David-Uraz}, {Alecian}, \& {MiMeS
  Collaboration}}]{2019MNRAS.489.5669P}
{Petit}, V., {Wade}, G.~A., {Schneider}, F.~R.~N., {et~al.} 2019, \mnras, 489,
  5669

\bibitem[{{Poe} {et~al.}(1990){Poe}, {Owocki}, \&
  {Castor}}]{1990ApJ...358..199P}
{Poe}, C.~H., {Owocki}, S.~P., \& {Castor}, J.~I. 1990, \apj, 358, 199

\bibitem[{{Powell} {et~al.}(1999){Powell}, {Roe}, {Linde}, {Gombosi}, \& {De
  Zeeuw}}]{1999JCoPh.154..284P}
{Powell}, K.~G., {Roe}, P.~L., {Linde}, T.~J., {Gombosi}, T.~I., \& {De Zeeuw},
  D.~L. 1999, Journal of Computational Physics, 154, 284

\bibitem[{{Puls} {et~al.}(2006){Puls}, {Markova}, {Scuderi}, {Stanghellini},
  {Taranova}, {Burnley}, \& {Howarth}}]{2006A&A...454..625P}
{Puls}, J., {Markova}, N., {Scuderi}, S., {et~al.} 2006, \aap, 454, 625

\bibitem[{{Puls} {et~al.}(2015){Puls}, {Sundqvist}, \&
  {Markova}}]{2015IAUS..307...25P}
{Puls}, J., {Sundqvist}, J.~O., \& {Markova}, N. 2015, in IAU Symposium, Vol.
  307, New Windows on Massive Stars, ed. G.~{Meynet}, C.~{Georgy}, J.~{Groh},
  \& P.~{Stee}, 25--36

\bibitem[{{Puls} {et~al.}(2008){Puls}, {Vink}, \&
  {Najarro}}]{2008A&ARv..16..209P}
{Puls}, J., {Vink}, J.~S., \& {Najarro}, F. 2008, \aapr, 16, 209

\bibitem[{{Rubio-D{\'\i}ez} {et~al.}(2021){Rubio-D{\'\i}ez}, {Sundqvist},
  {Najarro}, {Traficante}, {Puls}, {Calzoletti}, \&
  {Figer}}]{2021arXiv210811734R}
{Rubio-D{\'\i}ez}, M.~M., {Sundqvist}, J.~O., {Najarro}, F., {et~al.} 2021,
  arXiv e-prints, arXiv:2108.11734

\bibitem[{{Runacres} \& {Owocki}(2002)}]{2002A&A...381.1015R}
{Runacres}, M.~C. \& {Owocki}, S.~P. 2002, \aap, 381, 1015

\bibitem[{{Rybicki} {et~al.}(1990){Rybicki}, {Owocki}, \&
  {Castor}}]{1990ApJ...349..274R}
{Rybicki}, G.~B., {Owocki}, S.~P., \& {Castor}, J.~I. 1990, \apj, 349, 274

\bibitem[{{Schneider} {et~al.}(2020){Schneider}, {Ohlmann}, {Podsiadlowski},
  {R{\"o}pke}, {Balbus}, \& {Pakmor}}]{2020MNRAS.495.2796S}
{Schneider}, F.~R.~N., {Ohlmann}, S.~T., {Podsiadlowski}, P., {et~al.} 2020,
  \mnras, 495, 2796

\bibitem[{{Schneider} {et~al.}(2019){Schneider}, {Ohlmann}, {Podsiadlowski},
  {R{\"o}pke}, {Balbus}, {Pakmor}, \& {Springel}}]{2019Natur.574..211S}
{Schneider}, F. R.~N., {Ohlmann}, S.~T., {Podsiadlowski}, P., {et~al.} 2019,
  \nat, 574, 211

\bibitem[{{Schr{\o}der} {et~al.}(2021){Schr{\o}der}, {MacLeod}, {Ramirez-Ruiz},
  {Mandel}, {Fragos}, {Loeb}, \& {Everson}}]{2021arXiv210709675S}
{Schr{\o}der}, S.~L., {MacLeod}, M., {Ramirez-Ruiz}, E., {et~al.} 2021, arXiv
  e-prints, arXiv:2107.09675

\bibitem[{{Shore} \& {Brown}(1990)}]{1990ApJ...365..665S}
{Shore}, S.~N. \& {Brown}, D.~N. 1990, \apj, 365, 665

\bibitem[{{Shultz} {et~al.}(2020){Shultz}, {Owocki}, {Rivinius}, {Wade},
  {Neiner}, {Alecian}, {Kochukhov}, {Bohlender}, {ud-Doula}, {Landstreet},
  {Sikora}, {David-Uraz}, {Petit}, {Cerraho{\u{g}}lu}, {Fine}, {Henson}, {MiMeS
  Collaboration}, \& {BinaMIcS Collaboration}}]{2020MNRAS.499.5379S}
{Shultz}, M.~E., {Owocki}, S., {Rivinius}, T., {et~al.} 2020, \mnras, 499, 5379

\bibitem[{{Sobolev}(1960)}]{1960mes..book.....S}
{Sobolev}, V.~V. 1960, {Moving envelopes of stars} ({Harvard University Press})

\bibitem[{{Sudnik} \& {Henrichs}(2016)}]{2016A&A...594A..56S}
{Sudnik}, N.~P. \& {Henrichs}, H.~F. 2016, \aap, 594, A56

\bibitem[{{Sundqvist} {et~al.}(2019){Sundqvist}, {Bj{\"o}rklund}, {Puls}, \&
  {Najarro}}]{2019A&A...632A.126S}
{Sundqvist}, J.~O., {Bj{\"o}rklund}, R., {Puls}, J., \& {Najarro}, F. 2019,
  \aap, 632, A126

\bibitem[{{Sundqvist} \& {Owocki}(2013)}]{2013MNRAS.428.1837S}
{Sundqvist}, J.~O. \& {Owocki}, S.~P. 2013, \mnras, 428, 1837

\bibitem[{{Sundqvist} {et~al.}(2012{\natexlab{a}}){Sundqvist}, {Owocki},
  {Cohen}, {Leutenegger}, \& {Townsend}}]{2012MNRAS.420.1553S}
{Sundqvist}, J.~O., {Owocki}, S.~P., {Cohen}, D.~H., {Leutenegger}, M.~A., \&
  {Townsend}, R. H.~D. 2012{\natexlab{a}}, \mnras, 420, 1553

\bibitem[{{Sundqvist} {et~al.}(2018){Sundqvist}, {Owocki}, \&
  {Puls}}]{2018A&A...611A..17S}
{Sundqvist}, J.~O., {Owocki}, S.~P., \& {Puls}, J. 2018, \aap, 611, A17

\bibitem[{{Sundqvist} {et~al.}(2012{\natexlab{b}}){Sundqvist}, {ud-Doula},
  {Owocki}, {Townsend}, {Howarth}, \& {Wade}}]{2012MNRAS.423L..21S}
{Sundqvist}, J.~O., {ud-Doula}, A., {Owocki}, S.~P., {et~al.}
  2012{\natexlab{b}}, \mnras, 423, L21

\bibitem[{{Tanaka}(1994)}]{1994JCoPh.111..381T}
{Tanaka}, T. 1994, Journal of Computational Physics, 111, 381

\bibitem[{Thompson(1987)}]{THOMPSON19871}
Thompson, K.~W. 1987, Journal of Computational Physics, 68, 1

\bibitem[{{Townsend} \& {Owocki}(2005)}]{2005MNRAS.357..251T}
{Townsend}, R.~H.~D. \& {Owocki}, S.~P. 2005, \mnras, 357, 251

\bibitem[{{Townsend} {et~al.}(2007){Townsend}, {Owocki}, \&
  {ud-Doula}}]{2007MNRAS.382..139T}
{Townsend}, R.~H.~D., {Owocki}, S.~P., \& {ud-Doula}, A. 2007, \mnras, 382, 139

\bibitem[{{Trigilio} {et~al.}(2004){Trigilio}, {Leto}, {Umana}, {Leone}, \&
  {Buemi}}]{2004A&A...418..593T}
{Trigilio}, C., {Leto}, P., {Umana}, G., {Leone}, F., \& {Buemi}, C.~S. 2004,
  \aap, 418, 593

\bibitem[{{ud-Doula} \& {Owocki}(2002)}]{2002ApJ...576..413U}
{ud-Doula}, A. \& {Owocki}, S.~P. 2002, \apj, 576, 413

\bibitem[{{ud-Doula} {et~al.}(2008){ud-Doula}, {Owocki}, \&
  {Townsend}}]{2008MNRAS.385...97U}
{ud-Doula}, A., {Owocki}, S.~P., \& {Townsend}, R.~H.~D. 2008, \mnras, 385, 97

\bibitem[{{ud-Doula} {et~al.}(2009){ud-Doula}, {Owocki}, \&
  {Townsend}}]{2009MNRAS.392.1022U}
{ud-Doula}, A., {Owocki}, S.~P., \& {Townsend}, R.~H.~D. 2009, \mnras, 392,
  1022

\bibitem[{{ud-Doula} {et~al.}(2013){ud-Doula}, {Sundqvist}, {Owocki}, {Petit},
  \& {Townsend}}]{2013MNRAS.428.2723U}
{ud-Doula}, A., {Sundqvist}, J.~O., {Owocki}, S.~P., {Petit}, V., \&
  {Townsend}, R.~H.~D. 2013, \mnras, 428, 2723

\bibitem[{{ud-Doula} {et~al.}(2006){ud-Doula}, {Townsend}, \&
  {Owocki}}]{2006ApJ...640L.191U}
{ud-Doula}, A., {Townsend}, R. H.~D., \& {Owocki}, S.~P. 2006, \apjl, 640, L191

\bibitem[{{van der Velden}(2020)}]{2020JOSS....5.2004V}
{van der Velden}, E. 2020, The Journal of Open Source Software, 5, 2004

\bibitem[{{Vink} {et~al.}(2001){Vink}, {de Koter}, \&
  {Lamers}}]{2001A&A...369..574V}
{Vink}, J.~S., {de Koter}, A., \& {Lamers}, H.~J.~G.~L.~M. 2001, \aap, 369, 574

\bibitem[{{Wade} {et~al.}(2015){Wade}, {Barb{\'a}}, {Grunhut}, {Martins},
  {Petit}, {Sundqvist}, {Townsend}, {Walborn}, {Alecian}, {Alfaro},
  {Ma{\'{\i}}z Apell{\'a}niz}, {Arias}, {Gamen}, {Morrell}, {Naz{\'e}}, {Sota},
  {ud-Doula}, \& {MiMeS Collaboration}}]{2015MNRAS.447.2551W}
{Wade}, G.~A., {Barb{\'a}}, R.~H., {Grunhut}, J., {et~al.} 2015, \mnras, 447,
  2551

\bibitem[{{Wade} {et~al.}(2012){Wade}, {Grunhut}, {Gr{\"a}fener}, {Howarth},
  {Martins}, {Petit}, {Vink}, {Bagnulo}, {Folsom}, {Naz{\'e}}, {Walborn},
  {Townsend}, \& {Evans}}]{2012MNRAS.419.2459W}
{Wade}, G.~A., {Grunhut}, J., {Gr{\"a}fener}, G., {et~al.} 2012, \mnras, 419,
  2459

\bibitem[{{Wade} {et~al.}(2016){Wade}, {Neiner}, {Alecian}, {Grunhut}, {Petit},
  {Batz}, {Bohlender}, {Cohen}, {Henrichs}, {Kochukhov}, {Landstreet},
  {Manset}, {Martins}, {Mathis}, {Oksala}, {Owocki}, {Rivinius}, {Shultz},
  {Sundqvist}, {Townsend}, {ud-Doula}, {Bouret}, {Braithwaite}, {Briquet},
  {Carciofi}, {David-Uraz}, {Folsom}, {Fullerton}, {Leroy}, {Marcolino},
  {Moffat}, {Naz{\'e}}, {Louis}, {Auri{\`e}re}, {Bagnulo}, {Bailey},
  {Barb{\'a}}, {Blaz{\`e}re}, {B{\"o}hm}, {Catala}, {Donati}, {Ferrario},
  {Harrington}, {Howarth}, {Ignace}, {Kaper}, {L{\"u}ftinger}, {Prinja},
  {Vink}, {Weiss}, \& {Yakunin}}]{2016MNRAS.456....2W}
{Wade}, G.~A., {Neiner}, C., {Alecian}, E., {et~al.} 2016, \mnras, 456, 2

\bibitem[{{Xia} {et~al.}(2018){Xia}, {Teunissen}, {El Mellah}, {Chan{\'e}}, \&
  {Keppens}}]{2018ApJS..234...30X}
{Xia}, C., {Teunissen}, J., {El Mellah}, I., {Chan{\'e}}, E., \& {Keppens}, R.
  2018, \apjs, 234, 30

\end{thebibliography}
\bibliographystyle{aa} 

\begin{appendix}

\section{Boundary condition specifications}\label{sec:appendix}

We here further describe the adopted boundary conditions at the lower radial boundary (stellar surface) in our simulations. In this near star region the flow is inherently sub-magnetosonic leading to a non-linear MHD interaction with information propagation between the numerical grid (stellar wind) and the boundary (stellar surface) such that there is a priori no guarantee that the wind can relax into an asymptotic state.

An often used method to assign mathematically consistent boundary conditions to a problem is by using knowledge of the flow characteristics \citep{HEDSTROM1979222,THOMPSON19871}. Since for a (magnetic) radiation-driven flow such characteristics-based boundary conditions remain undeveloped, we choose a more approximate method of fixing a number of MHD variables $(\rho, \varv_r, \varv_\theta, B_r, B_\theta)$ in the ghost cells equal to the number of inward propagating MHD characteristics.\footnote{Namely the number of wave velocities from $\varv_r, \varv_r \pm a_\mathrm{slow}, \varv_r \pm \varv_\mathrm{Alf}, \varv_r \pm a_\mathrm{fast}$ that are greater than zero.} Imposing the same number of constraints on MHD variables as there are incoming characteristics ensures that the MHD equations are neither under-, nor over-specified.

The lower boundary density is fixed using information from the sonic point density. In this work we choose a lower boundary density $\rho_0$ a factor five larger than the sonic point density \citep{1988ApJ...335..914O}. The radial velocity $\varv_r$ is extrapolated into the first ghost cell $i$ of the lower boundary, allowing the wind to dynamically adjust, using a second-order constant extrapolation
\begin{equation}
\varv_r^i = \frac{1}{3} \left( 4\varv_r^{i+1} - \varv_r^{i+2} \right)
\end{equation}
and kept constant in the $n$ ghost cells below: $\varv_r^n =\varv_r^i, \forall n<i$.. Following this method allows the wind to dynamically adjust during the simulation to the overlying wind conditions.

The poloidal component of the velocity $\varv_\theta$ is fixed to enforce flow along a magnetic field line in the boundary. However, we have found it beneficial to make a distinction based on the magnetic confinement $\eta_\star$. For cases of no magnetic confinement $\eta_\star \leq 1$ we impose the condition $\varv_\theta^i=0$ in all ghost cells. In case of magnetic confinement, $\eta_\star>1$, a distinction is made between a `wind zone' and `dead zone' \citep{1968MNRAS.138..359M}, that is a transition happens at the colatitude where the last closed loop of the magnetosphere occurs:
\begin{equation}
\varv_\theta^i =
\begin{cases}
 \varv_r^i B_\theta^i/B_r^i & \text{if } | \pi/2 - \theta| > \arcsin \left( \sqrt{R_\star/R_A} \right), \\
 0                                                   &\text{elsewhere}.
\end{cases}
\end{equation}

This distinction ensures that in the wind zone near the poles, where $B_\theta$ can undergo significant deviations due to the wind outflow, the flow is still parallel to the magnetic field based on the constraints from the MHD induction equation. Within the magnetosphere we fix the poloidal velocity simply at zero.

The radial magnetic field component $\delta B_r = B_{\mathrm{tot},r} - B_{\mathrm{dip},r}$ is fixed using $r^{-2} \d{(r^2 B_{\mathrm{tot},r})}/\d{r} =0$ to ensure a dipole field is introduced at the lower boundary
\begin{equation}
\delta B_r^i = B_\star \left( \frac{R_\star}{r^i} \right)^2 + \delta B_{\star,r} - B_{\mathrm{dip},r}^i, \qquad \text{with } B_\star = B_p \cos \theta,
\end{equation}
while $\delta B_{\star,r}$ is computed in the code from solving the MHD equations with Tanaka's method and $B_{\mathrm{dip},r}^i$ follows from Eq.~\eqref{eq:initbfield}.

Lastly, the poloidal magnetic field component $\delta B_\theta$ is also extrapolated into the lower boundary using a second-order accurate constant extrapolation
\begin{equation}
\delta B_\theta^i = \frac{1}{3} \left( 4\delta B_\theta^{i+1} - \delta B_\theta^{i+2} \right).
\end{equation}
With these conditions adopted for the magnetic field we also ensure that the divergence-free constraint, Eq.~\eqref{eq:divb}, is satisfied outside the physical boundary of our problem.

\section{Comparison to MHD models of CAK winds}\label{sec:comparison}

\subsection{Radiation force modification and setup}

We compare the dynamical and morphological wind structures arising in our magnetic LDI models with those that assume a CAK line-force parametrisation. So far, in multi-dimensional numerical radiation-MHD simulations of line-driven winds the CAK line force $g_\mathrm{CAK}$ has been taken as purely radial and spherically symmetric \citep[e.g.][]{2002ApJ...576..413U,2017AN....338..868K,2019MNRAS.489.3251D}. As a result we run simulations with the line force contribution to the total radiation force in Eq.~\eqref{eq:mom} given by $\mathbf{g}_\mathrm{line} = g_\mathrm{CAK}\mathbf{\hat{e}_r}$ for which
\begin{equation}
g_\mathrm{CAK} = \frac{f_d(r)}{1-\alpha} \frac{(\kappa_e \bar{Q})^{1-\alpha} L_\star}{4\pi r^2 c^{\alpha+1}}  \left( \frac{1}{\rho} \frac{\d{\varv}}{\d{r}} \right)^\alpha,
\end{equation}
with $f_d(r)$ the one-dimensional finite-disc correction factor \citep{1986ApJ...311..701F,1986A&A...164...86P} to account for radiation coming from the full stellar disc. The other quantities appearing have the same meaning as in Sect.~\ref{sec:hydro}.

In performing simulations for a magnetic CAK wind, we employ the same basic simulation setup as described in Sect.~\ref{sec:hydro}. Here, however, we lower the amount of radial zones to $n_r=280$ (with $\Delta r_{i+1}/\Delta r_i = 1.02$) and colateral zones to $n_\theta=120$ because in the Sobolev approximation it is not computationally necessary to resolve small spatial scales. We also discuss here only a $\eta_\star=15$ model that has a confinement similar to several observations of magnetic O-stars \citep{2013MNRAS.429..398P}.

\subsection{Comparison}

\begin{figure*}[p]
\centering
\includegraphics[width=0.85\hsize]{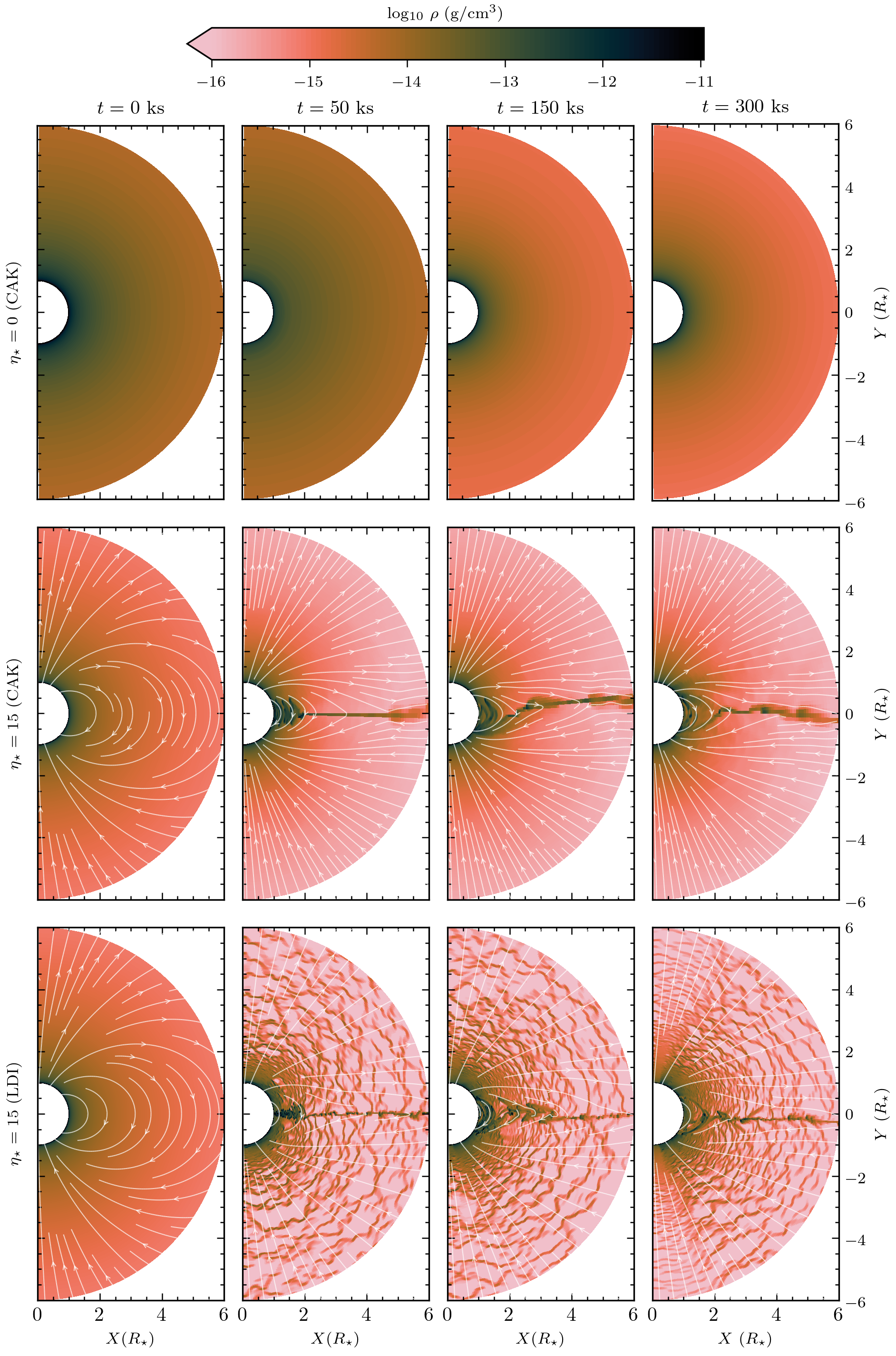}
\caption{Logarithmic wind density (in CGS units) displaying the evolution in time for the CAK and LDI models: (top) non-magnetic CAK wind with $\eta_\star=0$, (middle) strongly confined magnetic CAK wind with $\eta_\star=15$, and (bottom) strongly confined magnetic LDI wind with $\eta_\star=15$. The overplotted white lines in the magnetic wind models denote streamlines to illustrate the magnetic field topology.}
\label{fig:rhodyncak}
\end{figure*}

In analogy with Sect.~\ref{sec:rholdi} we make a direct comparison in Fig.~\ref{fig:rhodyncak} between a non-magnetic CAK model, a strong $\eta_\star=15$ CAK model, and the $\eta_\star=15$ LDI model discussed above. As in Fig.~\ref{fig:rhodyn} the wind density evolution is shown at the same timestamps. 

The non-magnetic CAK wind ($\eta_\star=0$) starts from a beta-velocity law
\begin{equation}\label{eq:betalaw}
\varv_r(r) = \varv_\infty \left( 1 - \frac{R_\star}{r} \right)^\beta,
\end{equation}
with $\beta=0.8$ appropriate for accounting for the finite stellar disk and $\varv_\infty$ the CAK terminal wind speed. When introducing a strong magnetic confinement, $\eta_\star=15$, in this CAK wind, the magnetic field lines at low latitudes remain closed thereby confining the wind flow \citep{2002ApJ...576..413U,2017AN....338..868K}. Part of this confined flow falls back towards the star, but some mass-leakage of overdense material occurs in the magnetic equatorial plane due to the compression of matter of colliding wind material from opposite footprints. Outside the magnetosphere, magnetic field lines with footprints at higher latitudes are progressively unable to counteract the wind flow and stretch into a near radial configuration. Within these regions the wind remains smooth and steady in close analogy with the non-magnetic CAK wind. However, this wind morphology is quite in contrast with what is seen in the strongly confined LDI wind, that is smooth flow vs.~wind sheets outside the magnetosphere. The occurrence of such wind sheets might then potentially also be found in observational signatures as discussed in Sec.~\ref{sec:obs}.

\end{appendix}

\end{document}